\documentclass[aps,prd,letter,twocolumn,superscriptaddress,floatfix,nofootinbib,showpacs,amsmath,amssymb,altaffilletter]{revtex4-1}

\usepackage{graphicx}
\usepackage{dcolumn}
\usepackage{bm}
\usepackage{color}
\usepackage{amsmath}
\usepackage[T1]{fontenc}
\usepackage{amsmath}
\usepackage{makecell}
\usepackage{braket}
\setcellgapes{2pt}
\usepackage{soul}

\newcommand{\BigO}[1]{\ensuremath{\operatorname{O}\bigl(#1\bigr)}}

\newcommand\blfootnote[1]{%
  \begingroup
  \renewcommand\thefootnote{}\footnote{#1}%
  \addtocounter{footnote}{-1}%
  \endgroup
}

\begin{document}


\title{Improved measurement of $^8$B solar neutrinos with 1.5 kt$\cdot$y of Borexino exposure}

\collaboration{Borexino Collaboration$^{*}$}\blfootnote{$^{*}$ spokesperson-borex@lngs.infn.it}

\newcommand{\Roma}{Present address: Dipartimento di Fisica, Sapienza Universit\`a di Roma e INFN, 00185 Roma, Italy}
\newcommand{\Napoli}{Present address: Dipartimento di Fisica, Universit\`a degli Studi Federico II e INFN, 80126 Napoli, Italy}
\newcommand{\Madrid}{Present address: Universidad Autónoma de Madrid, Ciudad Universitaria de Cantoblanco, 28049 Madrid, Spain}
\newcommand{\LNGSG}{Present address: INFN Laboratori Nazionali del Gran Sasso, 67010 Assergi (AQ), Italy}
\newcommand{\KFKI}{Also at: MTA-Wigner Research Centre for Physics, Department of Space Physics and Space Technology, Konkoly-Thege Miklós út 29-33, 1121 Budapest, Hungary}
\newcommand{\MPI}{Present address: Max-Planck-Institut f\"{u}r Physik, F\"{o}hringer Ring 6, 80805 M\"{u}nchen, Germany}
\newcommand{\Fermi}{Also at Fermilab National Accelerator Laboratory (FNAL), Batavia, IL 60510, USA}

\newcommand{\APC}{AstroParticule et Cosmologie, Universit\'e Paris Diderot, CNRS/IN2P3, CEA/IRFU, Observatoire de Paris, Sorbonne Paris Cit\'e, 75205 Paris Cedex 13, France}
\newcommand{\Dubna}{Joint Institute for Nuclear Research, 141980 Dubna, Russia}
\newcommand{\Genova}{Dipartimento di Fisica, Universit\`a degli Studi e INFN, 16146 Genova, Italy}
\newcommand{\Krakow}{M.~Smoluchowski Institute of Physics, Jagiellonian University, 30348 Krakow, Poland}
\newcommand{\Kiev}{Kiev Institute for Nuclear Research, 03680 Kiev, Ukraine}
\newcommand{\Kurchatov}{National Research Centre Kurchatov Institute, 123182 Moscow, Russia}
\newcommand{\Kurchatovb}{ National Research Nuclear University MEPhI (Moscow Engineering Physics Institute), 115409 Moscow, Russia}
\newcommand{\LNGS}{INFN Laboratori Nazionali del Gran Sasso, 67010 Assergi (AQ), Italy}
\newcommand{\Milano}{Dipartimento di Fisica, Universit\`a degli Studi e INFN, 20133 Milano, Italy}
\newcommand{\Perugia}{Dipartimento di Chimica, Biologia e Biotecnologie, Universit\`a degli Studi e INFN, 06123 Perugia, Italy}
\newcommand{\Peters}{St. Petersburg Nuclear Physics Institute NRC Kurchatov Institute, 188350 Gatchina, Russia}
\newcommand{\Princeton}{Physics Department, Princeton University, Princeton, NJ 08544, USA}
\newcommand{\PrincetonChemEng}{Chemical Engineering Department, Princeton University, Princeton, NJ 08544, USA}
\newcommand{\UMass}{Amherst Center for Fundamental Interactions and Physics Department, University of Massachusetts, Amherst, MA 01003, USA}
\newcommand{\Virginia}{Physics Department, Virginia Polytechnic Institute and State University, Blacksburg, VA 24061, USA}
\newcommand{\Munchen}{Physik-Department, Technische Universit\"at  M\"unchen, 85748 Garching, Germany}
\newcommand{\Lomonosov}{Lomonosov Moscow State University Skobeltsyn Institute of Nuclear Physics, 119234 Moscow, Russia}
\newcommand{\GSSI}{Gran Sasso Science Institute, 67100 L'Aquila, Italy}
\newcommand{\Dresda}{Department of Physics, Technische Universit\"at Dresden, 01062 Dresden, Germany}
\newcommand{\Mainz}{Institute of Physics and Excellence Cluster PRISMA+, Johannes Gutenberg-Universit\"at Mainz, 55099 Mainz, Germany}
\newcommand{\Juelich}{Institut f\"ur Kernphysik, Forschungszentrum J\"ulich, 52425 J\"ulich, Germany}
\newcommand{\RWTH}{III.  Physikalisches Institut B, RWTH Aachen University, 52062 Aachen, Germany}

\newcommand{\Tubingen}{Kepler Center for Astro and Particle Physics, Universit\"{a}t T\"{u}bingen, 72076 T\"{u}bingen, Germany}
\newcommand{\Huston}{Department of Physics, University of Houston, Houston, TX 77204, USA}
\newcommand{\Hamburg}{Institut f\"ur Experimentalphysik, Universit\"at Hamburg, 22761 Hamburg, Germany}

\author{M.~Agostini}\affiliation{\Munchen}
\author{K.~Altenm\"{u}ller}\affiliation{\Munchen}
\author{S.~Appel}\affiliation{\Munchen}
\author{V.~Atroshchenko}\affiliation{\Kurchatov}
\author{Z.~Bagdasarian}\affiliation{\Juelich}
\author{D.~Basilico}\affiliation{\Milano}
\author{G.~Bellini}\affiliation{\Milano}
\author{J.~Benziger}\affiliation{\PrincetonChemEng}
\author{D.~Bick}\affiliation{\Hamburg}
\author{D.~Bravo}\thanks{\Madrid}\affiliation{\Milano}
\author{B.~Caccianiga}\affiliation{\Milano}
\author{F.~Calaprice}\affiliation{\Princeton}
\author{A.~Caminata}\affiliation{\Genova}
\author{P.~Cavalcante}\thanks{\LNGSG}\affiliation{\Virginia}
\author{A.~Chepurnov}\affiliation{\Lomonosov}
\author{D.~D'Angelo}\affiliation{\Milano}
\author{S.~Davini}\affiliation{\Genova}
\author{A.~Derbin}\affiliation{\Peters}
\author{A.~Di Giacinto}\affiliation{\LNGS}
\author{V.~Di Marcello}\affiliation{\LNGS}
\author{X.F.~Ding}\affiliation{\Princeton}
\author{A.~Di Ludovico}\affiliation{\Princeton}
\author{L.~Di Noto}\affiliation{\Genova}
\author{I.~Drachnev}\affiliation{\Peters}
\author{A.~Formozov}\affiliation{\Dubna}\affiliation{\Milano}
\author{D.~Franco}\affiliation{\APC}
\author{C.~Galbiati}\affiliation{\Princeton}
\author{M.~Gschwender}\affiliation{\Tubingen}
\author{C.~Ghiano}\affiliation{\LNGS}
\author{M.~Giammarchi}\affiliation{\Milano}
\author{A.~Goretti\thanks{\LNGSG}}\affiliation{\Princeton}
\author{M.~Gromov}\affiliation{\Lomonosov}\affiliation{\Dubna}
\author{D.~Guffanti}\affiliation{\Mainz}
\author{C.~Hagner}\affiliation{\Hamburg}
\author{T.~Houdy}\thanks{\MPI}\affiliation{\APC}
\author{E.~Hungerford}\affiliation{\Huston}
\author{Aldo~Ianni}\affiliation{\LNGS}
\author{Andrea~Ianni}\affiliation{\Princeton}
\author{A.~Jany}\affiliation{\Krakow}
\author{D.~Jeschke}\affiliation{\Munchen}
\author{V.~Kobychev}\affiliation{\Kiev}
\author{G.~Korga}\thanks{\KFKI}\affiliation{\Huston}
\author{S.~Kumaran}\affiliation{\Juelich}\affiliation{\RWTH}
\author{T.~Lachenmaier}\affiliation{\Tubingen}
\author{M.~Laubenstein}\affiliation{\LNGS}
\author{E.~Litvinovich}\affiliation{\Kurchatov}\affiliation{\Kurchatovb}
\author{P.~Lombardi}\affiliation{\Milano}
\author{I.~Lomskaya}\affiliation{\Peters}
\author{L.~Ludhova}\affiliation{\Juelich}\affiliation{\RWTH}
\author{G.~Lukyanchenko}\affiliation{\Kurchatov}
\author{L.~Lukyanchenko}\affiliation{\Kurchatov}
\author{I.~Machulin}\affiliation{\Kurchatov}\affiliation{\Kurchatovb}

\author{S.~Marcocci$^{\dagger}$}\blfootnote{$^{\dagger}$ Deceased.}\thanks{\Fermi}
\affiliation{\GSSI}

\author{J.~Martyn}\affiliation{\Mainz}
\author{E.~Meroni}\affiliation{\Milano}
\author{M.~Meyer}\affiliation{\Dresda}
\author{L.~Miramonti}\affiliation{\Milano}
\author{M.~Misiaszek}\affiliation{\Krakow}
\author{V.~Muratova}\affiliation{\Peters}
\author{B.~Neumair}\affiliation{\Munchen}
\author{M.~Nieslony}\affiliation{\Mainz}
\author{R.~Nugmanov}\affiliation{\Kurchatov}
\author{L.~Oberauer}\affiliation{\Munchen}
\author{V.~Orekhov}\affiliation{\Mainz}
\author{F.~Ortica}\affiliation{\Perugia}
\author{M.~Pallavicini}\affiliation{\Genova}
\author{L.~Papp}\affiliation{\Munchen}
\author{\"O.~Penek}\affiliation{\Juelich}\affiliation{\RWTH}
\author{L.~Pietrofaccia}\affiliation{\Princeton}
\author{N.~Pilipenko}\affiliation{\Peters}
\author{A.~Pocar}\affiliation{\UMass}
\author{G.~Raikov}\affiliation{\Kurchatov}
\author{M.T.~Ranalli}\affiliation{\LNGS}
\author{G.~Ranucci}\affiliation{\Milano}
\author{A.~Razeto}\affiliation{\LNGS}
\author{A.~Re}\affiliation{\Milano}
\author{M.~Redchuk}\affiliation{\Juelich}\affiliation{\RWTH}
\author{A.~Romani}\affiliation{\Perugia}
\author{N.~Rossi}\thanks{\Roma}\affiliation{\LNGS}
\author{S.~Rottenanger}\affiliation{\Tubingen}
\author{S.~Sch\"onert}\affiliation{\Munchen}
\author{D.~Semenov}\affiliation{\Peters}
\author{M.~Skorokhvatov}\affiliation{\Kurchatov}\affiliation{\Kurchatovb}
\author{O.~Smirnov}\affiliation{\Dubna}
\author{A.~Sotnikov}\affiliation{\Dubna}
\author{Y.~Suvorov}\thanks{\Napoli}\affiliation{\LNGS}\affiliation{\Kurchatov}
\author{R.~Tartaglia}\affiliation{\LNGS}
\author{G.~Testera}\affiliation{\Genova}
\author{J.~Thurn}\affiliation{\Dresda}
\author{E.~Unzhakov}\affiliation{\Peters}
\author{A.~Vishneva}\affiliation{\Dubna}
\author{R.B.~Vogelaar}\affiliation{\Virginia}
\author{F.~von~Feilitzsch}\affiliation{\Munchen}
\author{M.~Wojcik}\affiliation{\Krakow}
\author{M.~Wurm}\affiliation{\Mainz}
\author{S.~Zavatarelli}\affiliation{\Genova}
\author{K.~Zuber}\affiliation{\Dresda}
\author{G.~Zuzel}\affiliation{\Krakow}


\begin{abstract}
We report on an improved measurement of the $^8$B solar neutrino interaction rate with the Borexino experiment at the Laboratori Nazionali del Gran Sasso. Neutrinos are detected via their elastic scattering on electrons in a large volume of liquid scintillator.
The measured rate of scattered electrons above 3 MeV of energy is $0.223\substack{+0.015 \\ -0.016}\,(stat)\,\substack{+0.006 \\ -0.006}\,(syst)$~cpd/100~t, which corresponds to an observed solar neutrino flux assuming no neutrino flavor conversion of $\Phi\substack{\rm ES \\ ^8\rm B}=2.57\substack{+0.17 \\ -0.18}(stat)\substack{+0.07\\ -0.07}(syst)\times$10$^6$~cm$^{-2}\,$s$^{-1}$. 
This measurement exploits the active volume of the detector in almost its entirety for the first time, and takes advantage of a reduced radioactive background following the 2011 scintillator purification campaign and of novel analysis tools providing a more precise  modeling of the background. Additionally, we set a new limit on the interaction rate of solar \textit{hep} neutrinos, searched via their elastic scattering on electrons as well as their neutral current-mediated inelastic scattering on carbon, $^{12}$C($\nu,\nu'$)$^{12}$C* ($E_{\gamma}$= 15.1~MeV).
\end{abstract}


\maketitle

\section{Introduction}
\label{sec:intro}
Borexino detects solar neutrinos with the lowest energy threshold of all solar neutrino experiments to date. Recently, Borexino has performed a new comprehensive measurement of pp-chain solar neutrinos \cite{Agostini:2018uly}, improving previous results on $pp$, $^7$Be, and $pep$ neutrino interaction rates~\cite{Arpesella:2008mt, Collaboration:2011nga, Bellini:2014uqa, Bellini:2013lnn},   $^8$B neutrinos above $\sim$3 MeV~\cite{Bellini:2008mr}, and  setting a limit on the neutrino flux produced by the $^3$He-proton fusion (\textit{hep}).
With its latest set of measurements, Borexino has provided a robust test of the Standard Solar Model (SSM) \cite{Vinyoles:2016djt} and more precisely probed the flavor conversion of solar neutrinos, described well by the Mikheyev-Smirnov-Wolfenstein (MSW) mechanism in matter coupled with neutrino mixing~\cite{Mikheev:1986gs,Wolfenstein:1977ue}. 

The Borexino analysis leading to the refined measurement of $pp$, $pep$, and $^7$Be interaction rates, the latter reaching a precision of better than 3\%, is extensively described in reference \cite{Agostini:2018uly}. 
Here we describe the improved measurement of the $^8$B solar neutrino interaction rate, performed with a greater than 10-fold increase of exposure with respect to an earlier measurement~\cite{Bellini:2008mr} and using an electron energy threshold of $\sim$3~MeV, the lowest to date. The increased exposure is due to a extended data acquisition period as well as a significantly enlarged fiducial mass that includes almost the entire Borexino 278 t scintillator target. 
Other notable improvements include the reduction of radioactive backgrounds following the 2011 scintillator purification campaign, effective at strongly reducing $^{208}$Tl contamination, and a new multi-variate analysis method to constrain cosmogenic $^{11}$Be contamination. Finally, we identified and included in the model a new source of background induced by radiogenic neutrons, which was not part of the previous analysis.  

The Borexino apparatus is briefly described in Sec.~\ref{sec:borexino}.
Sections~\ref{sec:response}, \ref{sec:data}, \ref{sec:backgrounds}, \ref{sec:analysis} present the detector response, data selection, backgrounds, analysis, and results of the $^8$B solar neutrino measurement, respectively.
Section~\ref{sec:hep} presents an improved search for \textit{hep} solar neutrino interaction rate.

\section{Experimental apparatus}
\label{sec:borexino}
Borexino is located underground in the Laboratori Nazionali del Gran Sasso (LNGS) in central Italy at a depth of 3800\,m.w.e.  
Neutrinos are detected via elastic scattering on electrons in a 278 t (nominally) organic liquid scintillator target. The scintillator consists of pseudocumene solvent (PC, 1,2,4-trimethylbenzene) doped with 1.5\,g/l of PPO (2,5-diphenyloxazole), a fluorescent dye.  
The liquid scintillator is contained in a 125\,$\mu$m-thick, spherical nylon vessel of 4.25\,m nominal radius. Scintillation light is detected by 2212 (nominal) ETL 9351 8" photomultiplier tubes (PMTs) uniformly mounted on a 13.7 m-diameter stainless steel sphere (SSS). 
Two concentric spherical buffer shells separate the active scintillator from the PMTs and SSS (323\,t and 567\,t, respectively). They are filled with PC doped with  dimethyl phthalate (DMP) to quench unwanted peripheral scintillation and shield the central active target from radiation emitted by the PMTs and SSS. The two PC buffers are separated by a second 125\,$\mu$m-thick nylon membrane that prevents the diffusion of radon emanated  by the PMTs and by the SSS into the central scintillator volume. Everything inside the SSS constitutes the inner detector (ID).

The SSS is surrounded by a domed, cylindrical water tank (18.0 m diameter, 16.9 m height), containing 2100\,t of ultra-pure water and serving as an additional absorber for external $\gamma$-rays and neutrons from the laboratory cavern. The water tank is equipped with  208 8''~PMTs, and run as a \v{C}erenkov detector (and veto) of cosmic muons (outer detector, OD \cite{Bellini:2011yd}).  
A complete description of the Borexino detector can be found in \cite{Alimonti:2008gc}.

\section{Detector response}
\label{sec:response}
The modeling of the Borexino detector response has steadily improved since the beginning of data-taking in 2007. Invaluable information has been provided by extensive calibration campaigns~\cite{Back:2012awa}. Moreover, the large body of data recorded over a decade has enabled extensive optimization of the Monte Carlo simulation of the detector~\cite{Agostini:2017aaa}. The level of understanding of the Borexino apparatus has enabled us to extend the $^{8}$B neutrino analysis to the entire scintillator target, an almost 3-fold mass increase from the previous measurement~\cite{Bellini:2008mr}.


The adopted analysis provides handles to reject most of the background components on an event-by-event basis via specific selection cuts (Section \ref{sec:data}). 
The radial distribution of the events surviving these cuts is fitted to discriminate bulk events occurring inside the scintillator volume (including solar neutrino events) from backgrounds originating outside the scintillator (see Sections \ref{sec:backgrounds} and \ref{sec:analysis}). 
No assumption is made on the neutrino energy spectrum, which allows us to test for any deviation from the MSW prediction. An accurate monitoring of the time evolution of the detector response is necessary. 
An important example is offered by the monitoring of a scintillator leak into the buffer region, started in April  2008, which caused the scintillator nylon vessel to deform over time. 
This effect is amplified by the mixing due to convective currents induced by temperature variations in the detector hall.  
Another important time-dependent effect to consider is the  loss of PMTs and the variation of their performance over the years.

To appropriately model time-dependent effects, we generate Monte Carlo simulated data sets 
on a weekly basis, which incorporate an exact map of operating PMT's and their performance parameters, such as gain and dark noise.  
\begin{figure}
\includegraphics[width=.9\linewidth]{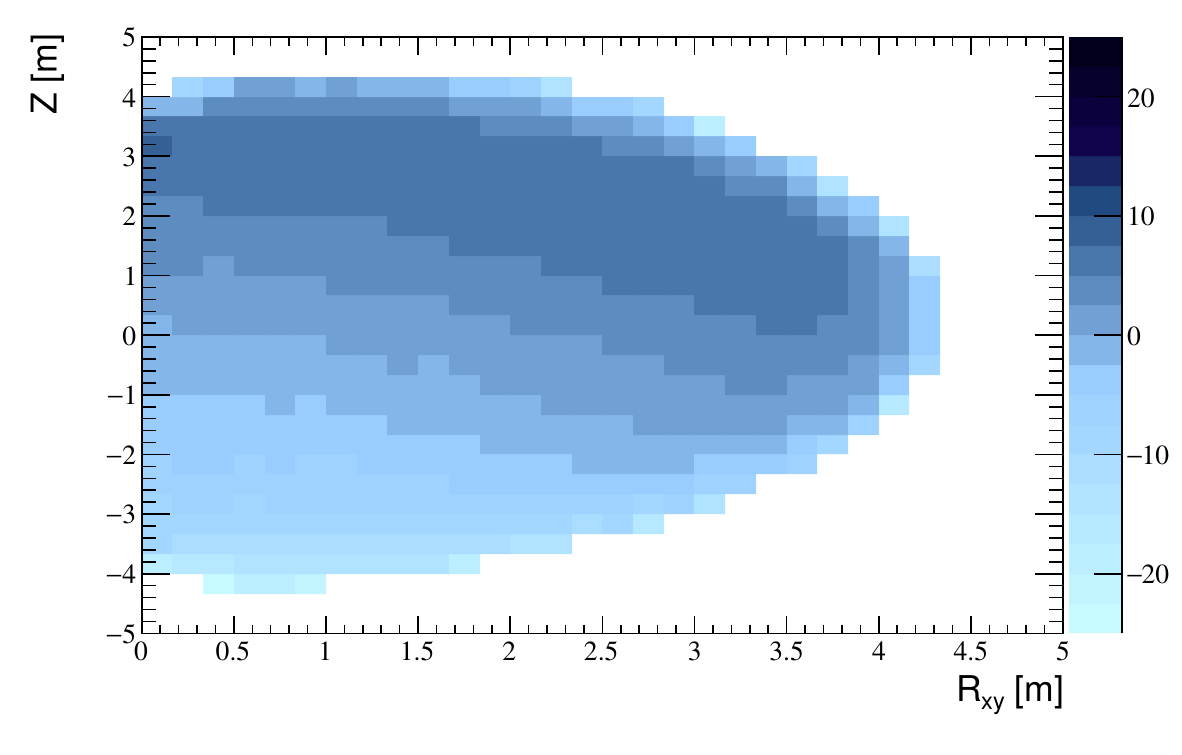}
\caption{\label{fig:emap} Relative  variation [\%], from Monte Carlo simulations, of the light yield with respect to the detector center, as function of the event reconstructed position (z vs R$_{xy}$=$\sqrt{x^2+y^2}$). The correspondent systematic error was estimated  in 1.9\%, using $^{241}$Am-$^9$Be calibration data. }
\end{figure}
In addition, the time-varying profile of the scintillator vessel shape is also included and updated every week. It is measured by locating background events generated by trace radioactive contamination embedded in the nylon film.
The  uncertainty on the reconstructed radial position of the nylon vessel is estimated at 1\% by comparing the reconstructed position of background events with the position of the membrane extracted from pictures taken with CCD cameras~\cite{Bellini:2013lnn, Back:2012awa}. 
Weekly simulated data sets contain 10$^2$--10$^3$ times events than the real data and are  collated after weighting them by the detector live time. 
This procedure applied to all the simulations used in this work, unless otherwise stated.

The energy threshold for this analysis is set at 3.2~MeV equivalent electron energy, with 50\% detection efficiency, to entirely reject 2.614~MeV $\gamma$-rays from $^{208}$Tl, due to $^{232}$Th contamination in the PMT's and the SSS. 
The energy calibration relies on the  characteristic $\gamma$ transitions from neutron captures on hydrogen and carbon, of 2.22~MeV and 4.95~MeV, respectively. 
Neutrons are emitted by a $^{241}$Am-$^9$Be source inserted in the scintillator. 
The light yield is defined as the sum of the integrated charge measured by each PMT and is expressed in photoelectrons (pe). In the central region of the detector (R$<$1~m) it is $\sim$500 pe/MeV/2000 PMT's: the Monte Carlo model reproduces it to within 1\% \cite{Back:2012awa,Agostini:2017aaa}. 

The non-uniformity of the spatial distribution of working PMTs, together with the scintillator light attenuation causes the energy response to depend on the event position.  
The Monte Carlo model predicts a relative variation of the light yield with respect to the center that ranges from -23\% in the bottom hemisphere to +8\% in the upper hemisphere (see Figure  \ref{fig:emap}).

The energy response of the model was validated by comparing data collected with the $^{241}$Am-$^9$Be calibration source with simulations: the relative difference of the light yield, $\Delta$LY, varies with radius up to a few \% at the edge of the scintillator volume. 
The associated uncertainty is 1.6\% when computed as the RMS of the $\Delta$LY distribution, weighted by the  event density. 
The overall error on the Monte Carlo energy response is equal to 1.9\%, which combines the uncertainties on the absolute light yield at the detector center and the relative, position-dependent variation. 

\begin{figure}
\includegraphics[width=.9\linewidth]{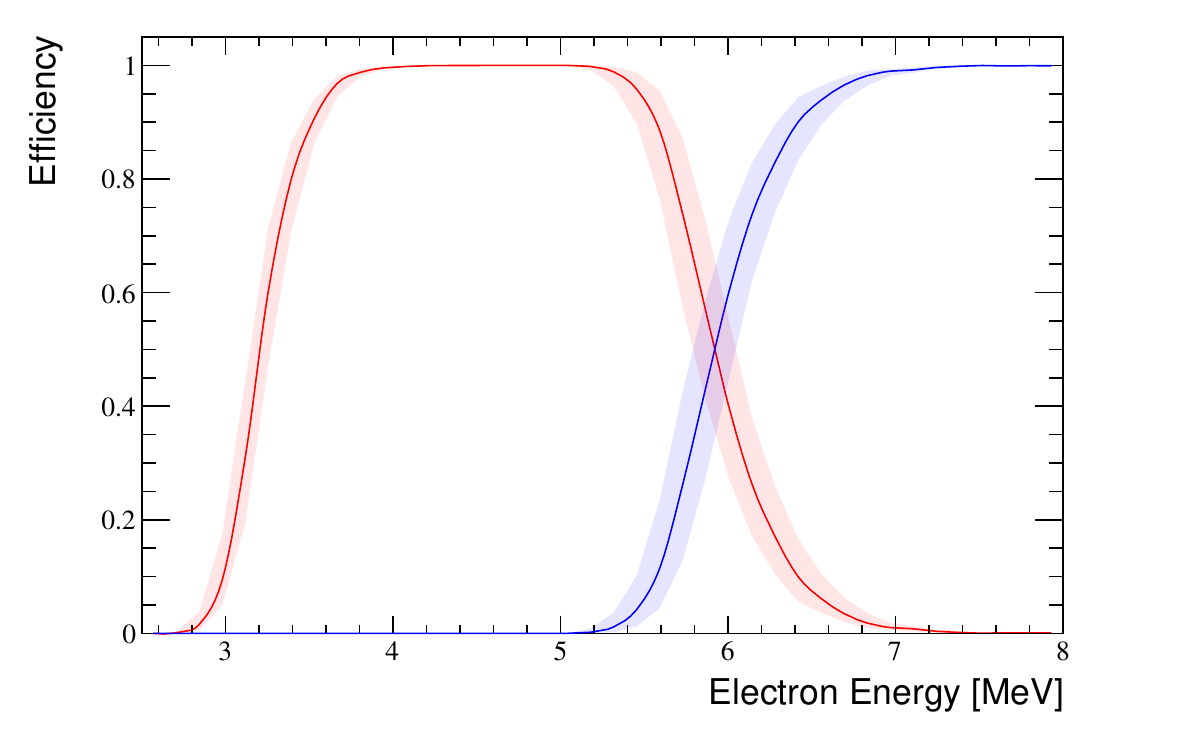}
\caption{\label{fig:thresholds} Detection efficiencies and associated uncertainties (due to the electron energy scale determination) of the HER-I ([1650, 2950] p.e., red line) and HER-II ([2950, 8500] p.e., blue line) ranges as a function of electron energy.  The detection efficiency equals 1 up to the 8500 p.e. upper edge ($\sim$17~MeV) of the HER-II range.}
\end{figure}

The energy threshold for the analysis is set at 1650 p.e., corresponding to 3.2~MeV electron energy, with detection efficiency of 50\% in the whole volume, as shown in Figure \ref{fig:thresholds}. 
The threshold is higher than the one used in our previous analysis (1494 p.e.) to take into account the higher light collection efficiency for events at high radius in the upper hemisphere, as illustrated in Figure \ref{fig:emap}. 
This region was previously excluded by a volume cut at 3~m radius.  
The upper limit is set at 8500 p.e. ($\sim$17~MeV electron energy), to fully accept $^8$B neutrino-induced recoil electrons.  

Radial fits to the energy spectra in two sub-ranges are independently performed. A lower energy range (HER-I),  with [1650, 2950] p.e., including events from natural radioactivity, and a higher energy range (HER-II), with [2950, 8500] p.e., dominated by external $\gamma$-rays following neutron capture processes on the SSS, as discussed in Sections \ref{sec:data} and \ref{sec:backgrounds}. 
In Borexino, the energy deposited by natural, long-lived radioactivity never exceeds 5~MeV (the Q-value of the $\beta$-decay of $^{208}$Tl), since the scintillation signal from $\alpha$'s of higher kinetic energy are quenched and fall below the analysis threshold.  
Of the residual backgrounds surviving selection cuts (see Section \ref{sec:data}), only cosmogenic $^{11}$Be and $\gamma$'s from neutron capture reactions make it into both energy windows.  
The signal detection efficiency associated to the energy cuts is evaluated by simulating neutrino events distributed uniformly throughout the active volume. The fractional number of events selected within the HER-I and HER-II ranges, converted to the recoil electron energy scale, is shown in Figure \ref{fig:thresholds}.

\section{Data selection}
\label{sec:data}
This work is based on data collected between January 2008 and December 2016 and  corresponds to 2062.4 live days of data, inclusive of the 388.6 live days of data used in the 2010 measurement. 
Data collected during detector operations such as scintillator purification and calibrations is omitted. 

Results from the HER-II sample use data from the entire active volume, while the HER-I sample requires a spatial cut to remove the top layer of scintillator. 
This is motivated by the presence of PPO from a scintillator leak, in proximity of the polar region, into the upper buffer fluid volume. Scintillation light from this buffer region have a chance to be mis-reconstructed at smaller radius and with energy at $\sim$3~MeV threshold. 
The z-cut to remove leak-related events was conservatively set at 2.5~m. The impact of this cut is investigated as a potential source of systematic uncertainty (see Section \ref{sec:analysis}). 


The active mass is evaluated with a toy Monte Carlo approach, by measuring the fraction of events falling within the scintillator volume for a set generated in a volume that includes it. 
The time-averaged mass is 266.0$\pm$5.3~ton, and assumes a scintillator density of 0.878$\pm$0.004~g/cm$^3$  \cite{Bellini:2013lnn}. 
The total exposure is 1,519~t$\cdot$y, a $\sim$11.5-fold increase with respect to our previous analysis. 
The mass fraction for the HER-I sample, after the z-cut at 2.5~m, is 0.857$\pm$0.006, obtained using a full optical simulation that includes effects from the spatial reconstruction of events. 

Data are selected with the following cuts, already discussed in reference \cite{Bellini:2008mr}: 
\begin{itemize}

\item \textit{muon cut}: events are rejected that either have more than 6 PMTs in the OD hit within 150 ns, or are identified by the ID as having a scintillation pulse mean time $>$100~ns or a peak time $>$30~ns;                                                                                                                                                                                                                                                                                                                                                                                                                                                                                                                                                                                                                                                                                                                                                                                                                                                                                                                                                                                                                                                                                                                                                                                                                                                                                                                                                                                                                                                                                                                                                                                                                                                                                                                                                                                                                                                                                                                                                                                                                                                                                                                                                                                                                                                                                                                                                                                                                                                                                                                                                                                                                                                                                                                                                                                                                                                                                                                                                                                                                                                                                                                                                                                                                                                                                                                                                                                                                                                                                                                                                                                                                                                                                                                                                                                                                                                                                                    

\item \textit{neutron cut}:  a 2 ms veto is applied after each muon detected by both ID and OD, to remove cosmogenic neutron captures on  $^{12}$C in the scintillator and in the buffer;

\item \textit{fast cosmogenics cut}:  a 6.5 s veto is applied after each muon crossing the scintillator to remove cosmogenic isotopes with life times between a few ms and 1.2 s ($^{12}$B, $^{8}$He, $^9$C, $^9$Li, $^8$B, $^6$He, and $^7$Li);

\item \textit{run start/break cut}:  a 6.5 s veto is applied at the beginning of each run to remove fast cosmogenic activity from muons missed during run restart;

\item $^{10}$C \textit{cut}:  a spherical volume of 0.8~m radius around all muon-induced neutron captures is vetoed for 120~s to reject cosmogenic $^{10}$C ($\tau$ = 27.8~s);

\item $^{214}$Bi-Po  \textit{cut}:  $^{214}$Bi and $^{214}$Po delayed coincident decays ($^{214}$Po-$\tau$ = 236 $\mu$s) are identified and rejected with $\sim$90\% efficiency. 

\end{itemize}

Muons crossing the water tank but not the SSS (\textit{external muons}) are detected by the OD with an efficiency >99.25\%~\cite{Bellini:2011yd}.
Muons crossing the scintillator (\textit{internal muons}) are defined either by using simultaneous signals from the ID and OD or by analysing the scintillation pulse shape in the ID alone. 
The pulse shape selection variables are the mean and peak times of the scintillation time profile. 
An event is identified as an internal muon if the mean time of the scintillation is $>$100~ns or the peak time is $>$30~ns. This cut introduces an inefficiency in the neutrino selection of 0.5\%, evaluated with Monte Carlo simulations. 
The rate of residual muons contaminating the HER-I and HER-II samples are measured following the procedure described in~\cite{Bellini:2008mr} as (1.2$\pm$0.1)$\times$10$^{-4}$ and (3.8$\pm$0.3)$\times$10$^{-4}$ cpd/100~ton, respectively.      

The definition of internal muons adopted by the \textit{fast cosmogenic cut} additionally requires E$>$400 p.e. ($\sim$0.8 MeV) in order to contain the dead time introduced by the \textit{fast cosmogenic cut} itself. 
The  energy cut has virtually no impact on the rejection efficiency of cosmogenic background, since it only removes muons that traverse 10-20~cm of buffer liquid and are hence far away from the scintillator target. 
The internal muon tagging inefficiency introduced by this tighter cut is <8$\times$10$^{-5}$ \cite{Bellini:2011yd}.  


The rate of 4.95~MeV $\gamma$-rays from cosmogenic neutron captures on carbon, surviving the \textit{neutron cut}, is (0.72$\pm$0.02)$\times$10$^{-4}$ cpd/100~t, with a  mean capture time of $\sim$254.5~$\mu$s \cite{Bellini:2011yd}, a cosmogenic neutron rate of 90.2$\pm$3.1 cpd/100~t and a carbon-to-hydrogen neutron cross section ratio of $\sim$1\% \cite{Bellini:2013pxa}.

The residual rate of fast cosmogenics after the \textit{fast cosmogenic cut} is (2.4$\pm$0.1)$\times$10$^{-3}$ cpd/100~t, obtained by fitting the distribution of elapsed time between each event and the previous muon (see~\cite{Bellini:2008mr} for details). 

The residual contamination of $^{10}$C surviving the $^{10}$C \textit{cut}, is evaluated as in \cite{Bellini:2008mr, Bellini:2013pxa}.  
The $^{10}$C \textit{cut} is effective only on  ``visible'' reaction  channels, {\it i.e.} those for which at least one neutron is emitted in association with $^{10}$C production. 
The selection efficiency is $0.925\substack{+0.075 \\ -0.200}$~\cite{Bellini:2013pxa} and the ``visible''  $^{10}$C rate is found to be $0.48\substack{+0.04 \\ -0.11}$ cpd/100~t, in good agreement with the previous measurement (0.52$\substack{+0.13 \\ -0.09}$~cpd/100~t \cite{Bellini:2013pxa}).  
The residual rate from visible channels is 4.8$\substack{+0.4 \\ -1.0}$$\times$10$^{-4}$ cpd/100~ton, which includes events surviving the $^{10}$C \textit{cut}, the energy cut, which rejects 98.3\% of $^{10}$C events, and the \textit{fast cosmogenic} cut, with an additional 17\% rejection efficiency.  
The residual rate from invisible production channels, dominated by the $^{12}$C(p,t)$^{10}$C reaction and evaluated in~\cite{Bellini:2008mr}, is (4.7$\pm$14.1)$\times$10$^{-4}$ cpd/100~t, after \textit{energy} and  \textit{fast cosmogenic} cuts.  

The $^{214}$Bi-Po \textit{cut} identifies $^{214}$Bi events correlated in time and space with the $^{214}$Po daughter nucleus. 
The closely-occurring events are searched in a [0.02, 1.4]~ms time window and with a maximum separation of 1~m. 
In addition, we require that the Gatti parameter, an $\alpha$/$\beta$ pulse shape discrimination estimator \cite{Back:2007gp}, for the $^{214}$Po to be >-0.008.  
The overall efficiency of this cut is 0.91~\cite{Bellini:2008mr}. 
The fraction of $^{214}$Bi with energy larger than the lower 1650 p.e. analysis threshold is derived directly from this rejected sample and is 6$\times$10$^{-4}$. 
The residual $^{214}$Bi rate leaking into the HER-I energy window is thus (2.2$\pm$1.0)$\times$10$^{-4}$ cpd/100~t.

\begin{table}[h]
\makegapedcells
\centering
\caption{\label{tab:residuals} Residual background rates, after selection cuts, in the HER-I [1650, 2950] p.e. and HER-II [2950, 8500] p.e. ranges, as discussed in Section \ref{sec:data}. The particular case of $^{11}$Be is discussed in Section \ref{sec:backgrounds}.}
\begin{ruledtabular}
\begin{tabular}{lcc}  
 Background    & HER-I rate &  HER-II    rate    \\ 
 &  [10$^{-4}$ cpd/100~t] &    [10$^{-4}$ cpd/100~t] \\ 
\hline
Fast cosmogenics &13.6$\pm$0.6 & 10.4$\pm$0.4 \\
Muons & 1.2$\pm$0.1 & 3.8$\pm$0.3\\
Neutrons  &0.72$\pm$0.02 &0 \\
$^{10}$C & 9.5$\pm$14.1& 0\\
$^{11}$Be & 0$\substack{+36.3 \\ -0.0}$ & 0$\substack{+54.9 \\ -0.0}$ \\
$^{214}$Bi & 2.2$\pm$1.0 &0 \\
\hline
  Total   & 27.2$\substack{+38.9 \\ -14.1}$ & 14.2$\substack{+54.9 \\ -0.5}$ \\ 
\end{tabular}
\end{ruledtabular}
\end{table}


The dead time introduced by all cuts is estimated with the toy Monte Carlo method. The selection cuts depend on internal and external muons and neutrons, so we generate artificial events with a constant rate (1.2 Hz) uniformly distributed in the IV and we add muon and neutron events selected from data with their time stamp.  The real vessel profiles are adopted for each week of data. After applying the  selection cuts to this hybrid dataset, we find the dead  time fraction to be 27.6\%. $^{214}$Bi-Po \textit{cut} is the only cut which does not depend on muons, however it does not introduce any relevant dead time due to the extremely low rate (about $\sim$70~$^{214}$Bi-Po candidates per day). After dead time subtraction, the residual detector livetime, is 1494 live days.


The rate of candidate events emerging from the selection cuts is 4.02 cpd. 
Untagged muons elude these cuts and could induce bursts of cosmogenic isotopes.
To suppress this source of background we require a minimum time difference of 5~s between events. 
The expected number of random coincidences in a 5~s window is 1.4 in the whole data set, corresponding to an additional dead time fraction of 2.5$\times$10$^{-4}$. 
A total of 17 events is rejected by this cut. 

\begin{figure}
\includegraphics[width=.9\linewidth]{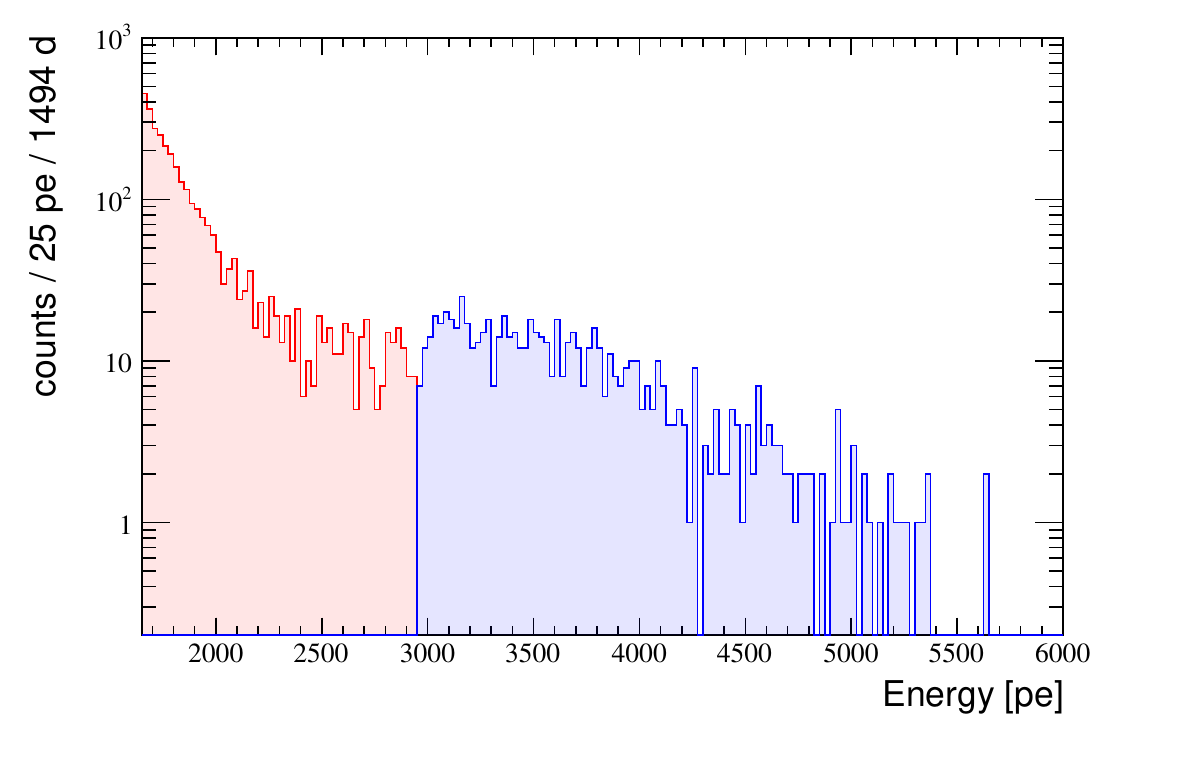}
\caption{\label{fig:energy} Energy spectrum of residual events after selection cuts in the HER-I (red) and HER-II (blue) ranges.  A z-cut at 2.5~m is applied to the HER-II region (see text) causing the step in the number of selected events visible at 2950 pe.  No events are found in the HER-II sub-range [6000, 8500] p.e..}
\end{figure}

The final sample comprises of 6065 candidate events surviving all selection cuts, with an exposure of 1089$\pm$21~t$\cdot$y after dead time subtraction. 
The resulting energy spectrum is shown in Figure \ref{fig:energy} for both HER-I and HER-II energy windows.
Residual background rates, after selection cuts, are listed in table \ref{tab:residuals}.

\section{Untagged backgrounds}
\label{sec:backgrounds}
In this section we report on strategies developed to characterize backgrounds that survive selection cuts and cannot be identified on an event-by-event basis. 
Four sources of background are of this kind. 
Two of them, bulk $^{208}$Tl and \textit{in situ} produced cosmogenic $^{11}$Be, are uniformly distributed within the scintillator volume.
Another is represented by decays of $^{208}$Tl embedded in the nylon vessel or on its surface, and the last, {\it i.e.} high energy $\gamma$-rays from neutron captures on peripheral detector components, are external to the IV. 

\subsection{$^{208}$Tl contamination}\label{sec:tl208}
$^{208}$Tl is produced by the decay of $^{212}$Bi with 36\% branching ratio. It is a $\beta$-decay with Q-value = 5.0 MeV, simultaneously emitting an electron and $\gamma$ rays. 
The $^{208}$Tl activity is quantified by looking at the alternative $^{212}$Bi decay mode (64\% BR) and counting the $^{212}$Bi-$^{212}$Po delayed coincidences.
The short mean life ($\tau$ = 431 ns) of $^{212}$Po together with the space correlation between the two decays, makes these coincidences easy to identify with little to no background, allowing for an accurate estimation of $^{212}$Bi, and hence of $^{208}$Tl.  
For the latter we measure an activity of (1.8$\pm$0.3)$\times$10$^{-2}$ cpd/100~t. 
This is about 5 times lower than in earlier work~\cite{Bellini:2008mr}, a consequence of the scintillator purification campaign occurred after the former search.  

\begin{figure}
\includegraphics[width=.9\linewidth]{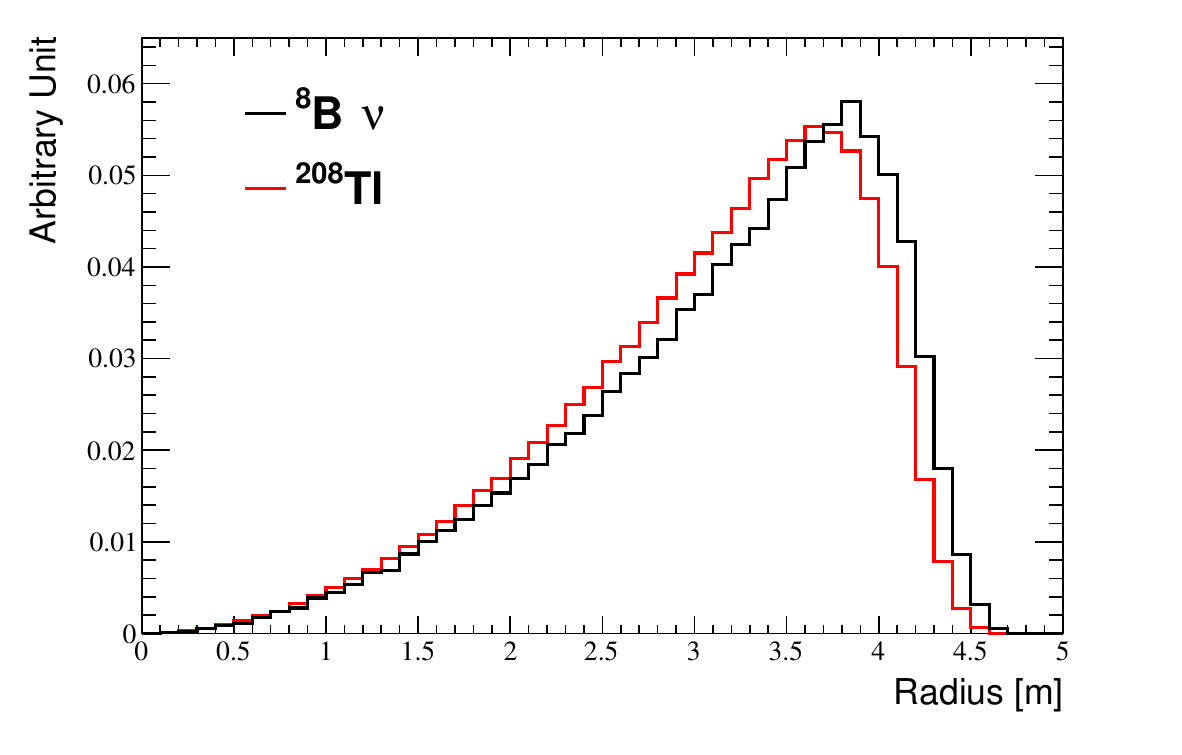}
\caption{\label{fig:tl208radial} Radial distributions of simulated $^8$B $\nu$ events (black) and $^{208}$Tl decays (red), with [1650, 2950] p.e. energy cut. The two spectra are normalized to the number of reconstructed events.  The difference is due to $^{208}$Tl emission $\gamma$-rays, escaping the scintillator.}
\end{figure}

In the outermost shell of scintillator, 2.6 MeV $\gamma$-rays from $^{208}$Tl decays (99\% BR) may escape into the buffer and shift the reconstructed event energy below the analysis threshold. This artificially shifts the radial distribution towards lower radii with respect to the neutrino induced electron recoils, as shown in Figure \ref{fig:tl208radial}. For this reason, $^{208}$Tl is included as a separate component in the radial fit, carrying a penalty factor derived from the uncertainty on its measured bulk activity. 
This represents a difference with respect to the previous analysis, where $^{208}$Tl background was statistically subtracted from the total event rate. 

\subsection{Cosmogenic $^{11}$Be}\label{sec:be11}
$^{11}$Be is a $\beta$ emitter with $Q=11.5$~MeV and $\tau\sim20$~s. In liquid scintillator, it is a product of muon spallation on $^{12}$C.  
\textit{In situ} $^{11}$Be production in liquid scintillator was observed by KamLAND at the Kamioka mine \cite{PhysRevC.81.025807}, where the mean muon energy $\langle E_{\mu} \rangle\sim$~260~GeV. 
On the contrary, both Borexino (at Gran Sasso $\langle E_{\mu} \rangle\sim$~280~GeV) and the NA54 experiment at CERN, which investigated the production rate of radioactive isotopes in liquid scintillator with a muon beam (at 100 and 190~GeV)~\cite{HAGNER200033}, were only able to set upper limits on the production of $^{11}$Be.  

Our previous $^8$B analysis relied on the extrapolation of the $^{11}$Be rate from the KamLAND measurement, yielding (3.2$\pm$0.6)$\times$10$^{-2}$ cpd/100~t above 3~MeV. The new Borexino estimation of the $^{11}$Be rate is based on a larger exposure and on a multivariate fit that includes the energy spectrum and the time profile of events with respect to the preceding muon. 

Candidate $^{11}$Be events are selected with $E>6$~MeV from a [10, 150]~s time window following the preceding muon. 
The lower time cut is set at 10~s to exclude events from other cosmogenic isotopes. 
To contain accidental background, the energy deposited by preceding muons must be larger than 5~MeV, and the $^{11}$Be candidate must be spatially confined within a 2~m radius from the muon track. 
The efficiency of the latter cut is $>$91.4\%, obtained by assuming that all $^{11}$Be isotopes are produced by neutron spallation via $^{12}$C(n,2p)$^{11}$Be reaction by neutrons with an average lateral distance of 81.4~cm from the muon track, as measured in~\cite{Bellini:2013pxa}.   
The assumption is conservative when considering that neutrons have the longest range of all muon-induced secondaries responsible of $^{11}$Be production, such as $\pi^-$, via $^{12}$C($\pi^-$,p)$^{11}$Be \cite{Galbiati:2004wx}, and $^7$Li, via $^{12}$C($^7$Li,$^8$B)$^{11}$Be.


Energy and  time difference,  with respect to the preceding muon, of  $^{11}$Be  candidates are simultaneously fitted in a multivariate mode with  two component models combining  $^{11}$Be signal, from  Monte Carlo simulations, and accidental background.  The latter is extracted directly from data, by  collecting events in the [150, 300]~s window after a muon and occurring more than 2~m away from the muon track. The fit  prefers a negative $^{11}$Be rate, and is compatible with zero.

When setting a boundary condition requiring only null or positive rates, the fit results a $^{11}$Be rate above 3 MeV of zero with a positive sigma of 9.1$\times$10$^{-3}$ cpd/100~t.  This is $\sim$3~$\sigma$ lower than the rate extrapolated from the KamLAND measurement. The observed low  rate   may be due to the  \textit{cosmogenic} and $^{10}$C cuts, which  affects also $^{11}$Be events. 
The $^{11}$Be rates in the HER-I and HER-II ranges are listed in Table \ref{tab:residuals}.

\subsection{Surface contamination}\label{sec:surface}

The nylon of the IV is the only material in direct contact with the scintillator, in addition to the plumbing of the filling and purification system. The IV was designed and constructed to make it as radio-pure as possible. 
The measured $^{238}$U and $^{232}$Th concentration in nylon is 5 and 20~ppt,  respectively \cite{Arpesella:2001iz}.
Nonetheless, while one of the cleanest solid materials ever assayed at the time Borexino was commissioned, its intrinsic radioactivity still exceeds that of the scintillator by many orders of magnitude. Events from the nylon vessel thus contribute to the event rate in the outermost shells of scintillator and call for volume fiducialization for most analyses.


\begin{figure}
\includegraphics[width=.9\linewidth]{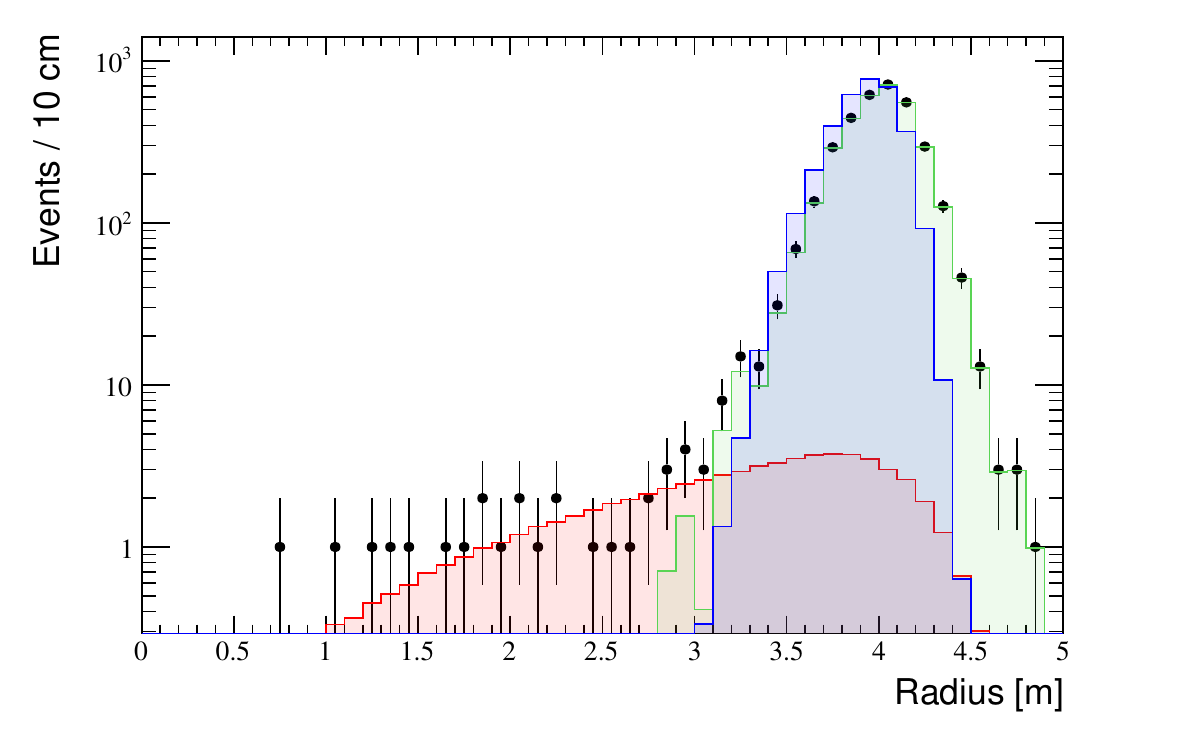}
\caption{\label{fig:bi212} Radial distribution of $^{212}$Bi events occurring in the scintillator, selected with the fast $^{212}$Bi-Po coincidences, with no energy degradation for the $\alpha$ decay (black dots). The distribution is fitted with a bulk (red) component and one from emanation and diffusion from the nylon vessel (green). The radial distribution of  $^{208}$Tl (blue) emanated from the vessel is derived from the latter, as described in the text. }
\end{figure}

The only vessel background causing events in the energy range of interest for this analysis is $^{208}$Tl, a daughter of $^{232}$Th embedded in the material. 
When it decays, $^{208}$Tl may find itself within the nylon membrane (contributing to what we call \textit{surface} events), or in the fluid in close proximity of the vessel. 
Two mechanism can cause $^{208}$Tl to leave the nylon. 
Firstly, nuclei in the $^{232}$Th decay chain may recoil into the liquid as a result of one of the intermediate decays. Alternatively, $^{220}$Rn, a volatile progenitor of $^{208}$Tl can diffuse out of the nylon into the scintillator during its 56~s half life.
Surface and \textit{emanation} components display different spatial distributions, which we model independently. 

Surface events are simulated by generating $^{208}$Tl decays uniformly across the nylon membrane.  
The emanated component cannot be reliably modeled because of the uncertainty introduced by convective motions of the scintillator~\cite{Bravo-Berguno:2017b,Bravo-Berguno:2017wqe}. 
The time delay between the appearance of $^{220}$Rn and the $^{208}$Tl decay is dominated by $^{212}$Pb ($\tau\sim$15~hr). 
The diffusion scale of $^{208}$Tl over this time interval suggests that it may decay several cm from the vessel surface, a visible effect in Borexino. 

The radial distribution of emanated $^{208}$Tl events is derived from the measured distribution of $^{212}$Bi-$^{212}$Po fast coincidences. 
When $\alpha$'s are emitted from $^{212}$Po implanted into the vessel, they lose a fraction of their energy inside the nylon and their scintillation signal appears degraded.
Surface events can thus be discarded by selecting $^{212}$Po decays with full $\alpha$ energy deposition in scintillator. 
To extract the $^{212}$Bi emanation component from the distribution of $^{212}$Bi-$^{212}$Po events, bulk $^{212}$Bi events are simulated and subtracted from the $^{212}$Bi data sample, after being normalized by their intrinsic contamination measured in the scintillator (see Figure \ref{fig:bi212}).




Despite $^{212}$Bi and $^{208}$Tl being equally located in space, their distributions differ because of energy-dependent resolution effects.
We derive the true "emanated" $^{212}$Bi radial distribution using the ROOT TSpectrum \textit{deconvolution} algorithm \cite{Brun:1997pa} to deconvolve the detector response. 
The difference between reconstructed and true radius for events generated 1~cm away from the vessel inside the scintillator is obtained from Monte Carlo simulation.
The true radial distribution is in turn convolved with the Monte Carlo spatial response function generated for $^{208}$Tl events, with the same procedure used for  $^{212}$Bi. 
The final $^{208}$Tl radial distribution is shown in Figure~\ref{fig:bi212}. 

The surface and emanation radial distributions of $^{208}$Tl are included in the fitting strategy, described in Section \ref{sec:analysis}. 
Uncertainties on the detector spatial response  for both $^{212}$Bi and  $^{208}$Tl events are included in the evaluation of systematic uncertainties, as discussed in Section \ref{sec:analysis}. 

\subsection{Radiogenic neutron captures}\label{sec:neutron}

\begin{figure}[t]
\includegraphics[width=.9\linewidth]{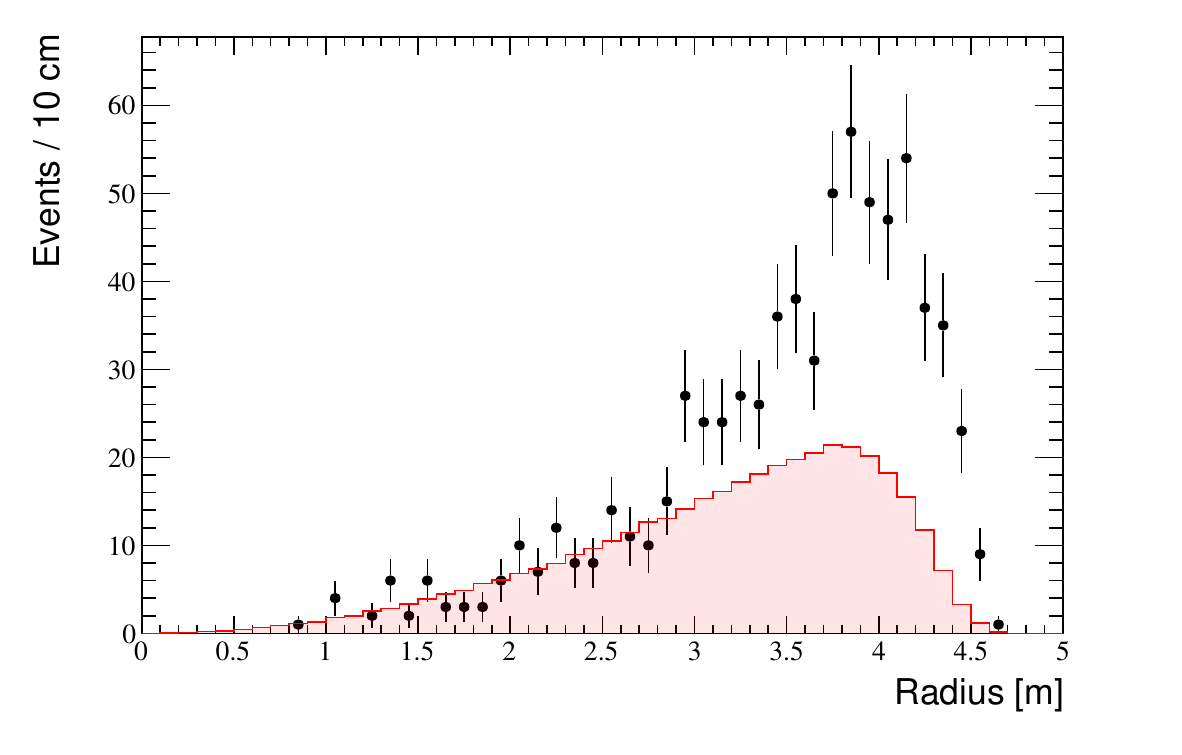}
\caption{\label{fig:externals} Radial distribution of $^8$B candidates (black dots) compared with the Monte Carlo (red) one for events with Q$>$2950 pe ($\sim$5MeV).  The excess at large radii is an indication for the additional high energy external background component induced by radiogenic neutron captures on detector materials.}
\end{figure}

\begin{figure}[t]
\includegraphics[width=.9\linewidth]{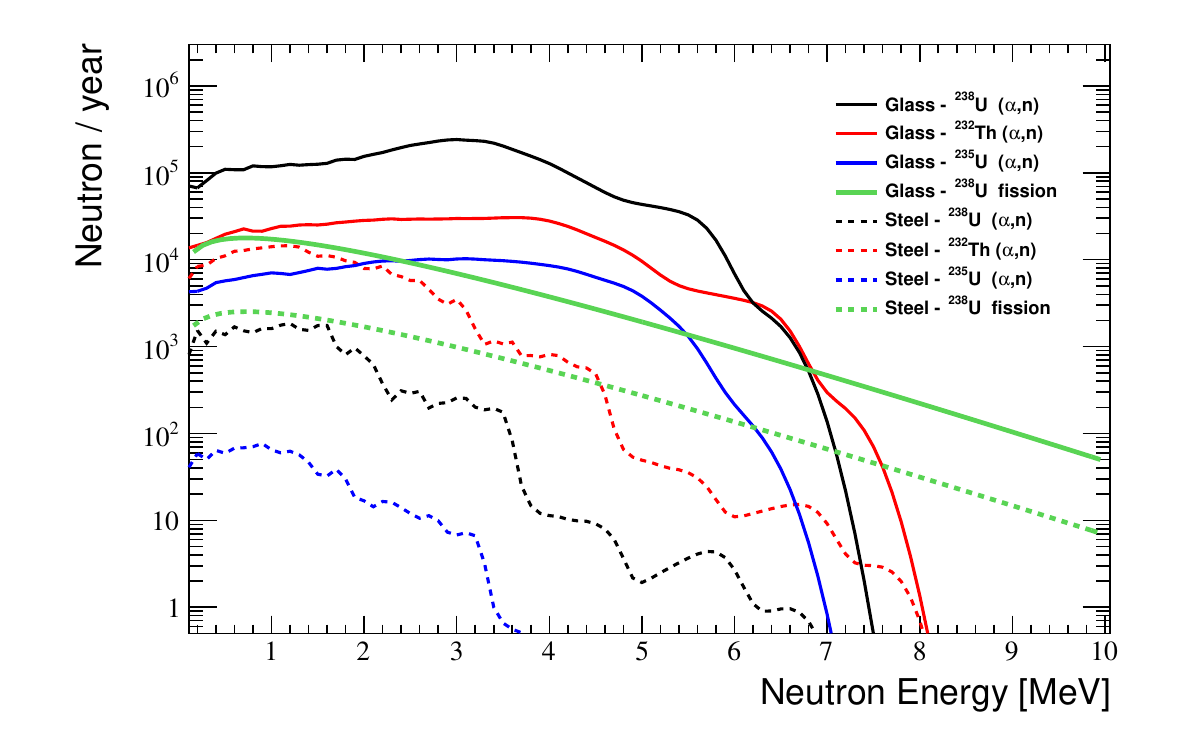}
\caption{\label{fig:neutrons} Predicted energy spectra and fluxes of neutrons produced via ($\alpha$, n) reactions and spontaneous fissions,  from  $^{238}$U, $^{235}$U, and $^{232}$Th contaminations in SSS and in PMT glasses. }
\end{figure}

\begin{table*}[t]
\makegapedcells
\centering
\caption{\label{tab:neutrons} Neutron fluxes from ($\alpha$, n) reactions and fissions from $^{238}$U,   $^{235}$U, and  $^{232}$Th  contaminations in stainless steel and PMT glass, as measured by the Borexino collaboration \cite{Arpesella:2001iz}.}
\begin{ruledtabular}
\begin{tabular}{l|ccc|ccc}
    & \multicolumn{3}{c|}{SSS (45~t)}  & \multicolumn{3}{c}{PMT Glass (1.77~t)}  \\
    &   $^{238}$U &   $^{235}$U &  $^{232}$Th   &   $^{238}$U &   $^{235}$U &  $^{232}$Th \\ 
\hline
 Concentration [g/g]  \cite{Arpesella:2001iz}&3.7 10$^{-10}$ & 2.7 10$^{-12}$& 2.8 10$^{-9}$ &  6.6 10$^{-8}$& 4.8 10$^{-10}$ & 3.2 10$^{-8}$ \\   
\hline
 ($\alpha$, n) rate [n/decay]  \cite{TALYS} &5.0 10$^{-7}$ & 3.8 10$^{-7}$& 1.9 10$^{-6}$ &  1.6 10$^{-5}$& 1.9 10$^{-5}$ & 1.8 10$^{-5}$ \\   
 ($\alpha$, n) neutron flux [year$^{-1}$] &3.3 10$^{3}$ & 1.2 10$^{2}$& 3.1 10$^{4}$ &  7.3 10$^{5}$& 4.1 10$^{4}$ & 1.3 10$^{5}$ \\   
 \hline
Spontaneous fission rate [n/(g s)]\cite{fission} & 1.36 10$^{-2}$ & 3.0 10$^{-4}$  & $<$1.32 10$^{-7}$ &   1.36 10$^{-2}$ & 3.0 10$^{-4}$  & $<$1.32 10$^{-7}$ \\   
Spontaneous fission neutron flux [year$^{-1}$] &7.1 10$^{3}$ & $\BigO{<10}$ & $\BigO{<1}$ &  5.0 10$^{4}$& $\BigO{<10}$ &$\BigO{<1}$ \\   
\end{tabular}
\end{ruledtabular}
\end{table*}

The HER-II data sample should only contain bulk scintillator events, namely $^8$B neutrinos and cosmogenic $^{11}$Be. 
No contributions from long-lived, natural radioactivity with E$>$5~MeV is expected. 
However, the data show an excess of events at large radii at odds with a bulk distribution, as shown in Figure \ref{fig:externals}. 
This effect was not previously observed because of limited statistics in the fiducial volume within 3~m radius. 

The excess can be explained by $\gamma$-rays arising from the capture of radiogenic neutrons produced in detector materials via ($\alpha$, n) or spontaneous fission reactions.  
Two sufficiently massive detector components with non-negligible $^{238}$U, $^{235}$U, and $^{232}$Th contamination are identified as possible sources of neutrons: the $\sim$45~t SSS, and the glass of the PMT's (totaling $\sim$0.8 kg $\times$ 2212 PMT's $\sim$1.77~t). 

The $^{238}$U and $^{232}$Th contamination of the SSS and the PMT glass measured by the Borexino~\cite{Arpesella:2001iz} are reproduced in Table \ref{tab:neutrons}.  
The $^{235}$U contamination is obtained from $^{238}$U, imposing the natural isotopic ratio, and the neutron yield used in this study assumes secular equilibrium along each decay chain.  
Some Borexino detector components, such as the PMT dynode structure, are known to have a higher specific radioactive contamination, but are neglected here in light of their limited total mass.  

The mean number of ($\alpha$, n) neutrons per decay and their associated energy spectra in each material are evaluated for each decay chain with NEUCBOT~\cite{Westerdale:2017kml}, a tool based on the TALYS simulation package~\cite{TALYS} that also accounts for any $\alpha$ energy lost inside materials.   

The energy spectrum of neutrons produced by spontaneous fission reactions is modeled with the Watt's semi-empirical relation~\cite{PhysRev.89.1288}, as

\begin{equation}
\label{eq:watt}
f(E) \propto \textrm{Sinh} (\sqrt{2 E}) \, e^{-E},
\end{equation}
\noindent where E is in MeV. 
The corresponding neutron fluxes are derived from reference \cite{fission} and quoted in Table \ref{tab:neutrons}. 
 

Neutrons emerging from the SSS and the PMT glass are simulated with the Borexino Monte Carlo framework \cite{Agostini:2017aaa}, using the input energy distributions described above (Figure~\ref{fig:neutrons} and Eq.~\ref{eq:watt}). 
The simulation indicates that neutrons capture mainly on the iron of the SSS and on the hydrogen and carbon in the buffer fluid within $\sim$80~cm of the SSS, producing $\gamma$-rays with energies up to $\sim$10~MeV.    
These $\gamma$-rays are attenuated by the remaining $\sim$2~m-thick buffer fluid separating it from the scintillator volume. 
The fraction of events with $E>1650$~p.e. ranges between 10$^{-5}$ and 10$^{-4}$, depending on the location of the neutron capture and the energy of the emitted $\gamma$-ray. 
The simulations of energetic $\gamma$-rays from the detector periphery were validated in Borexino with the deployment of a $^{228}$Th calibration source~\cite{Agostini:2017aaa}.

\begin{figure}
\includegraphics[width=.9\linewidth]{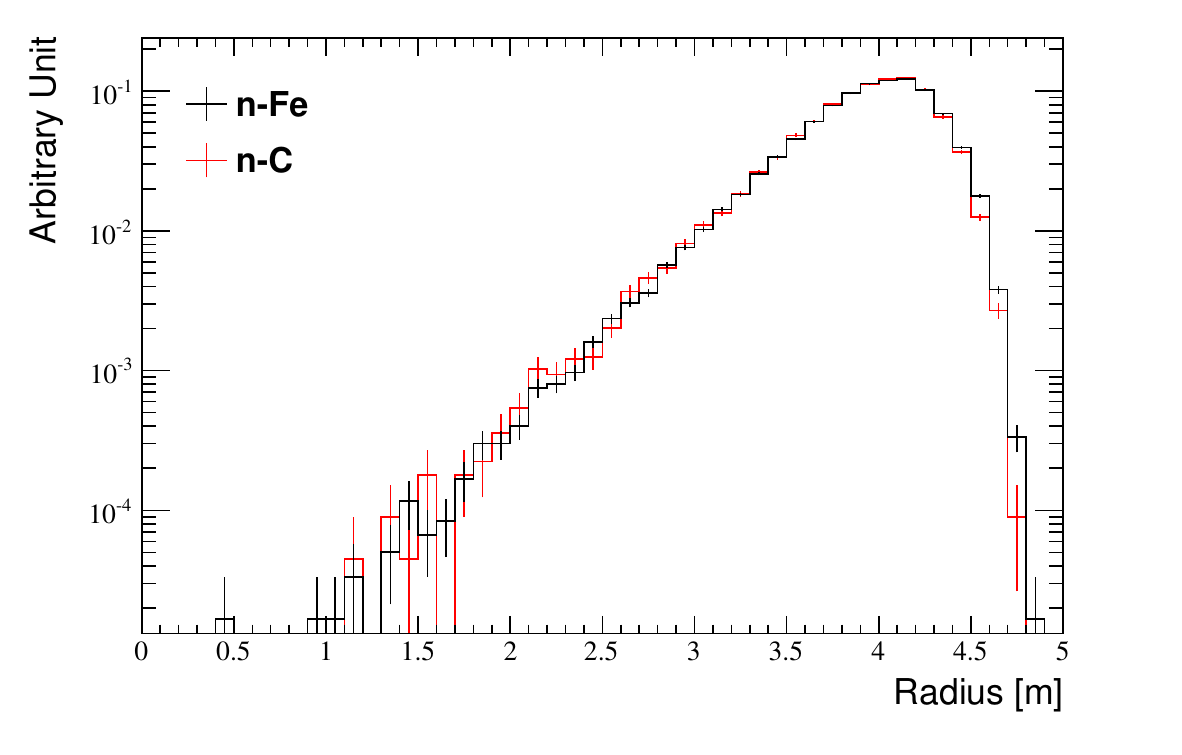}
\caption{\label{fig:extshapes} Reconstructed radial distributions of  simulated events with energy falling in  the HER-I range and generated by neutron captures on carbon in the buffer (red), and on iron in the SSS (black).  }
\end{figure}

We calculate that 148 (151) events are induced by neutrons in the HER-I (HER-II) data samples. 
The uncertainty on this estimation is dominated by the ($\alpha$, n) cross sections. 
As discussed in~\cite{Westerdale:2017kml}, disagreement of up to 100\% exists between data compilations and predictions by TALYS and SOURCES-4C~\cite{SHORES200178}, an alternative code for calculating $\alpha$-induced neutron fluxes.  
In the same reference, good agreement between the predicted neutron energy spectra is reported. 

We finally address possible systematic effects on the reconstructed radial distribution of events in the HER-I range. 
These arise from a possible imbalance between $\gamma$-rays produced in the buffer region and in the SSS and PMTs caused by a simplified model of the detector used in the simulations, which lacks certain details such as, \textit{e.g.}, the internal PMT structure and the cable feedthroughs.
We compare the distribution of events from neutron captures in the buffer and in the SSS and observe minor differences limited to the vessel edge, as shown in Figure \ref{fig:extshapes}. The impact of this systematic is evaluated in the next section.

\section{Data analysis}
\label{sec:analysis}
The identification of the neutrino signal relies on its different radial distribution from that of background.
Neutrino candidates are expected to be uniformly distributed throughout the scintillator, a property shared with $^{11}$Be background that is described by the same radial function but whose rate is, however, constrained as illustrated in Section~\ref{sec:backgrounds}.
The \textit{bulk} $^{208}$Tl on the contrary, follows a different distribution, as shown in Figure~\ref{fig:tl208radial} and discussed in Section~\ref{sec:backgrounds}.

\begin{figure}
\includegraphics[width=.9\linewidth]{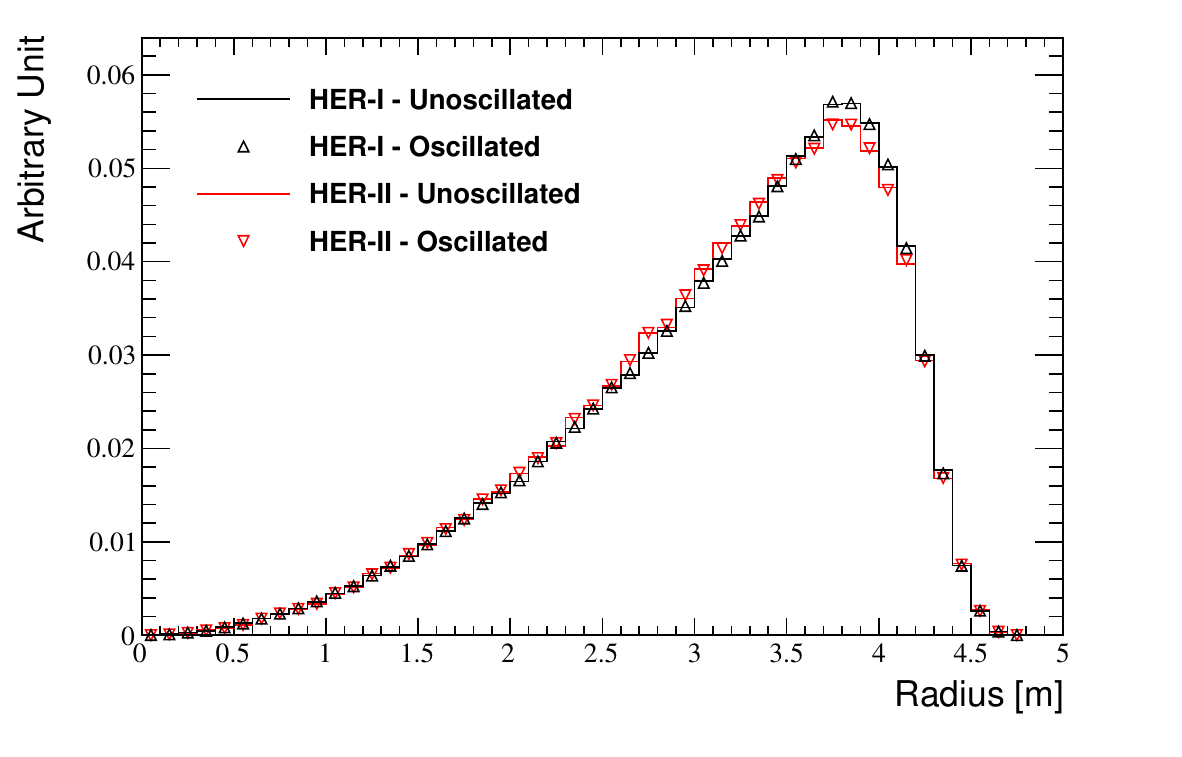}
\caption{\label{fig:intshapes} Comparison of the radial distributions from unoscillated and oscillated neutrinos in the HER-I and HER-II ranges. Oscillated and unoscillated spectra are almost indistinguishable, confirming the negligible dependence of the radial distribution on the energy spectrum. The MSW-LMA parameters used for simulating oscillating neutrinos are $\Delta$m$^2$=7.50$\times$10$^{-5}$~eV$^2$, sin$^2$($\theta_{12}$)=0.306, and sin$^2$($\theta_{13}$)=0.02166~\cite{Esteban:2016qun}.
}
\end{figure}

The $^8$B energy spectrum used here is that from W. Winter \textit{et al.}~\cite{PhysRevC.73.025503}.
Spectral distortions due to neutrino flavor conversion have no impact on the shape of the radial HER-I and HER-II distributions, as illustrated in Figure \ref{fig:intshapes}, where radial shapes simulated for both energy windows with and without MSW-LMA flavor conversion are compared. 

The radial fit estimator is the binned likelihood ratio, to account for empty bins at small radii. We include a penalty factor constraining the bulk $^{208}$Tl component to the known uncertainty on its rate. 

\begin{table}
\makegapedcells
\centering
\caption{\label{tab:fit} HER-I and HER-II rates of  signal and background components from the radial distribution fits (only statistical errors are quoted). Bulk events are dominated by $^8$B neutrinos, with contributions from $^{11}$Be decays and residual background quoted in Tab. \ref{tab:residuals} }
\begin{ruledtabular}
\begin{tabular}{lcc}  
 Component    & HER-I rate &  HER-II    rate    \\ 
    & [cpd/227.8~t] &  [cpd/266.0~t]    \\ 
\hline
Bulk events & 0.310$\pm$0.029& 0.235$\pm$0.021 \\
External & 0.224$\pm$0.078& 0.239$\pm$0.022\\
$^{208}$Tl bulk  & 0.042$\pm$0.008& - \\
$^{208}$Tl emanation & 0.469$\pm$0.063& - \\
$^{208}$Tl surface & 1.090$\pm$0.046&  - \\
\end{tabular}
\end{ruledtabular}
\end{table}

The HER-II data sample is fitted with 2 components only: $^8$B neutrinos and the external component from neutron captures. 
The HER-I sample requires three additional fit components, all due to $^{208}$Tl: \textit{bulk} (dissolved in the scintillator), \textit{surface} (intrinsic to the nylon vessel), and \textit{emanation} (diffused from the nylon vessel into the outer edge of scintillator).
The fit results are summarized in Table \ref{tab:fit}.

\begin{figure}
\includegraphics[width=.9\linewidth]{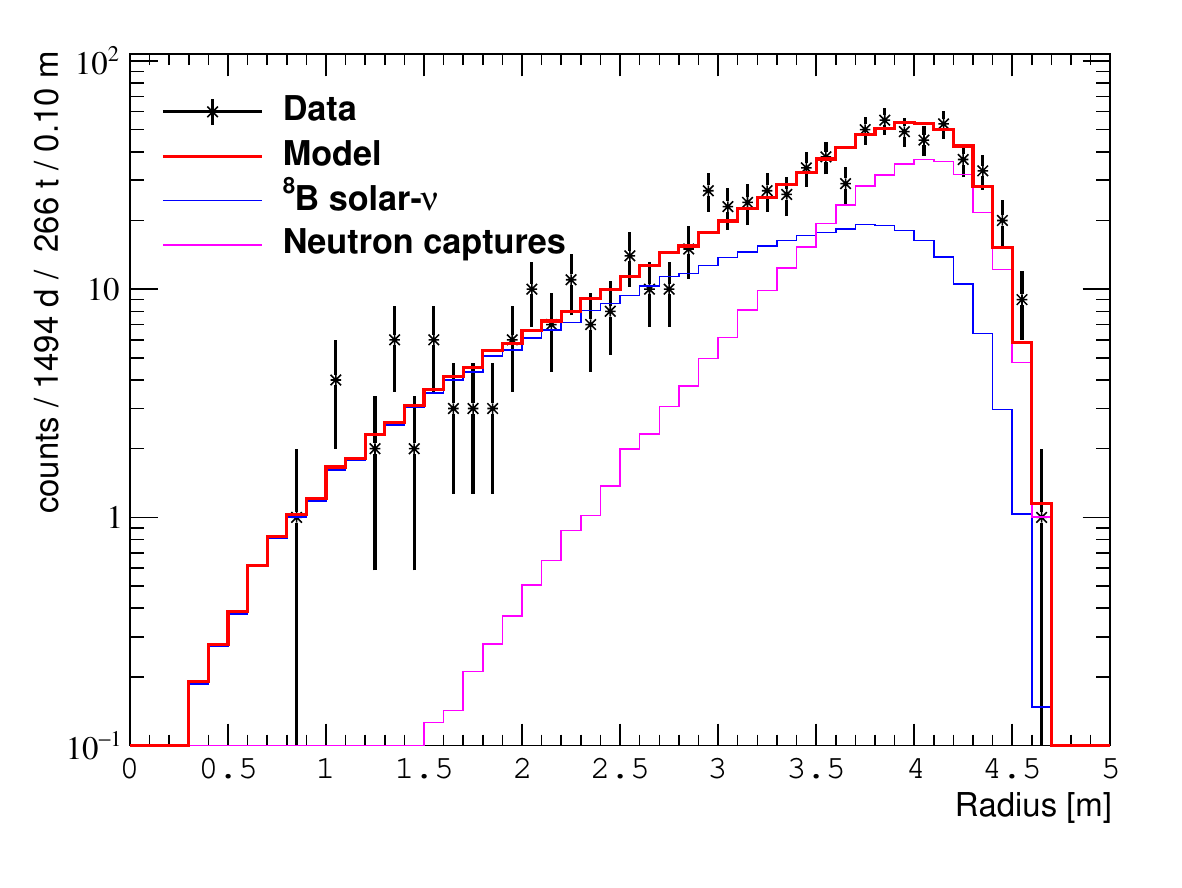}
\caption{\label{fig:fitHE} Fit of the event radial distribution in the HER-II range, [2950, 8500] p.e.. }
\end{figure}

\begin{figure}
\includegraphics[width=.9\linewidth]{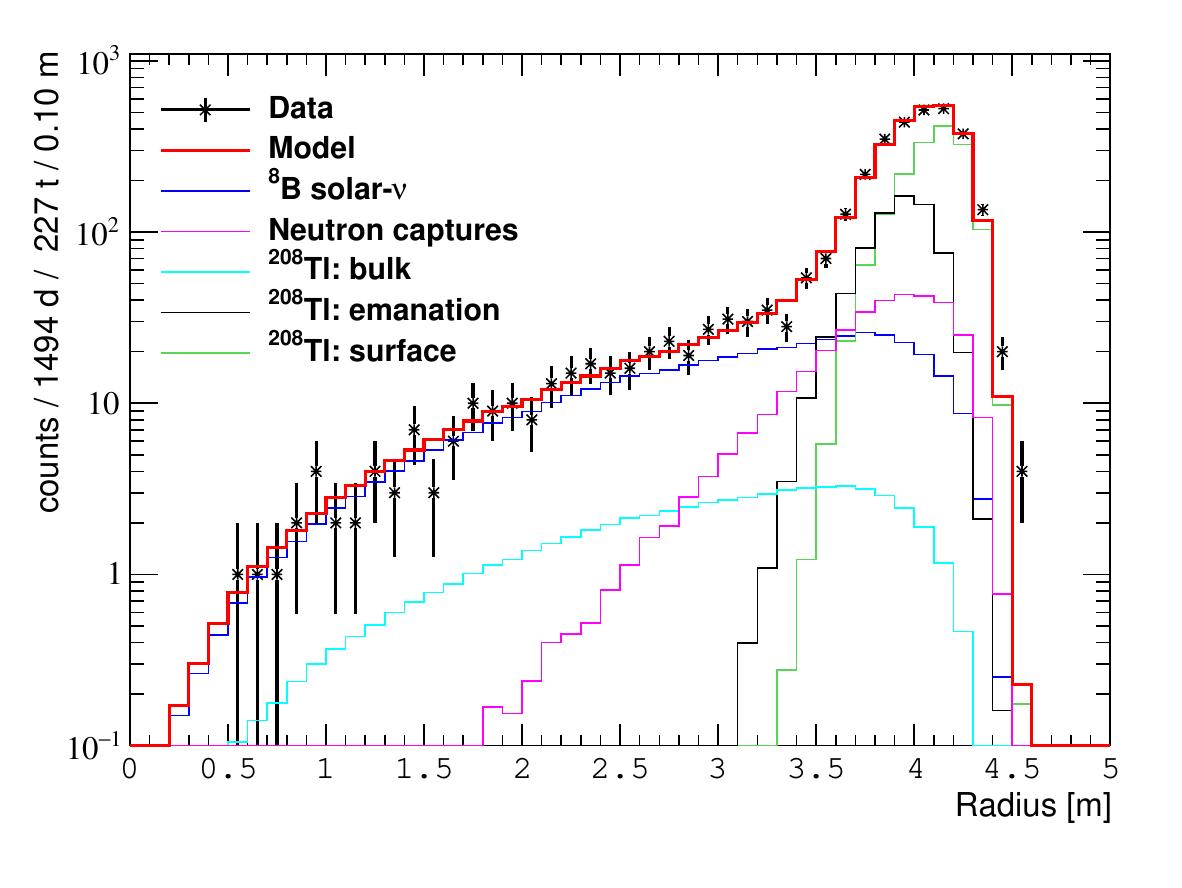}
\caption{\label{fig:fitLE} Fit of the event radial distribution in the HER-I range, [1650, 2950] p.e..  }
\end{figure}

The HER-II and HER-I radial fits are shown in Figure~\ref{fig:fitHE} and Figure~\ref{fig:fitLE}, and the corresponding $\chi^2$/dof, excluding empty bins, of 30.4/35 (HER-II) and 31.3/36 (HER-I), respectively. 
The \textit{emanation} $^{208}$Tl rate is measured at 0.47$\pm$0.06 cpd.  
It is worth mentioning that its exclusion from the LE fit leads to a $\chi^2$/dof of 91.6/36.  

The number of external neutron capture-induced events from the fit is 351$\pm$31 and 335$\pm$117 for the HER-II and HER-I ranges, respectively. 
In both cases the best-fit number is $\sim$2 times larger than predicted by simulations, still within less than 2$\sigma$, including model uncertainties. 

The best-fit normalization of the \textit{bulk} $^{208}$Tl component for the HER-I dataset is close to the central value expected from the rate of $^{212}$Bi-Po coincidences.  
The weak anti-correlation coefficient ($-0.299$) between $^8$B neutrinos and  $^{208}$Tl substantiates the ability of the fit to discriminate between these two distributions.  

The best-fit rate of $^8$B neutrinos, after subtraction of residual backgrounds itemized in Table \ref{tab:fit}, is 0.136$\pm$0.013 cpd/100~t for the HER-I energy range and 0.087$\substack{+0.008 \\ -0.010}$ cpd/100~t for the HER-II window. 
The total rate above 1650 p.e. is 0.223$\substack{+0.015 \\ -0.016}$~cpd/100~t. 


The result from the fit is stable (within 1$\sigma_{stat}$) to changes of the histogram binning and to a $\pm$3\% linear distortion of the simulated radius. A slight decrease in the normalized $\chi^2$ was observed by multiplying the simulated radius by 1.015, which improves the agreement at large radii. However, such a variation is small enough to induce any systematics in the radial fit. 


The fitted $^8$B neutrino interaction rates were tested to be stable to changes of the response function used for de-convolving (convolving) the $^{212}$Bi ($^{208}$Tl) spatial distribution, determining the radial profile of the \textit{emanation} $^{208}$Tl component (see Figure~\ref{fig:bi212}). 
Its stability was specifically tested with a response function simulating events located 6~cm away from the IV, inside the scintillator, and no appreciable effect was observed.

Finally, we tested the fit stability against variations of the radial shape of the neutron capture $\gamma$-rays component, assuming the limiting cases of neutrons exclusively capturing on the SSS or the buffer fluid, shown in Figure~\ref{fig:extshapes}. 
A smaller normalized $\chi^2$ is obtained when considering neutron captures on SSS only, but the $^8$B neutrino rate is stable within statistical uncertainty. 

The  systematic sources mostly affecting the result are the determination of the active mass and the uncertainty on the energy scale (both discussed in section \ref{sec:data}), and the z-cut applied in the HER-I range.   
To quantify the effect of the latter, we performed the fit with a modified z-cut, $\pm$0.5~m around the chosen value (2.5~m). 
The other systematic effects were evaluated with Monte Carlo simulations. 
Subdominant sources of systematic uncertainty relate to the scintillator density and to the live time estimation. 
Systematic uncertainties for the HER-I and HER-II ranges are collected in Table~\ref{tab:systematics}.

\begin{table}[h]
\makegapedcells
\centering
\caption{\label{tab:systematics} Systematic sources and percentage uncertainties of the measured rates in the HER-I, HER-I, and HER=HER-I+HER-II ranges.}
\begin{ruledtabular}
\begin{tabular}{lccc}
    & HER-I  & HER-II & HER  \\
   Source  & $\sigma$  &    $\sigma$  & $\sigma$      \\ 
\hline
   Active mass                  & 2.0  & 2.0 & 2.0\\
   Energy scale                & 0.5  & 4.9 &  1.7 \\
   z-cut                             & 0.7  & 0.0 & 0.4 \\
   Live time                      & 0.05 & 0.05 & 0.05 \\
   Scintillator density        & 0.05  &  0.05 & 0.05 \\
\hline
  Total                              & 2.2 & 5.3 &  2.7 \\ 
\end{tabular}
\end{ruledtabular}
\end{table}

In summary, the final $^8$B solar neutrino rates, corrected by the data selection efficiency,  in the HER-I, HER-II, and combined energy regions (HER=HER-I+HER-II) are:   
\begin{align*}
R_{HER-I}  &=& 0.136\substack{+0.013 \\ -0.013}\,(stat)\,\substack{+0.003 \\ -0.003}\,(syst)\;\mathrm{cpd/100\,t},  \\[2mm]
R_{HER-II} &=&  0.087\substack{+0.08 \\ -0.010}\,(stat)\,\substack{+0.005 \\ -0.005}\,(syst)\;\mathrm{cpd/100\,t},  \\[2mm]
R_{HER}  &=&  0.223\substack{+0.015 \\ -0.016}\,(stat)\,\substack{+0.006 \\ -0.006}\,(syst)\;\mathrm{cpd/100\,t}.
\end{align*}

The precision on the HER $^8$B rate measurement is $\sim$8\%, improved by more than a factor 2 with respect to our previous result~\cite{Bellini:2008mr}. 

The equivalent flavor-stable $^8$B neutrino flux inferred from this measurement is 2.57$\substack{+0.17 \\ -0.18}(stat)\substack{+0.07 \\ -0.07}(syst)$$\times$10$^6$\;cm$^{-2}$s$^{-1}$, in good agreement with the previous Borexino result of 2.4$\pm$0.4$\times$10$^6$\;cm$^{-2}$s$^{-1}$~\cite{Bellini:2008mr} and with the high-precision measurement by SuperKamiokande, 2.345$\pm$0.014$(stat)\pm$0.036$(syst)$$\times$10$^6$~cm$^{-2}$~s$^{-1}$ \cite{Abe:2016nxk}).

The expected $^8$B solar neutrino flux according to the B16 SSM~\cite{Vinyoles:2016djt} with high metallicity (GS98~\cite{Grevesse1998}) is 5.46$\pm$0.66$\times$10$^6$\;cm$^{-2}$s$^{-1}$. 
The apparently missing flux is fully compatible with neutrino flavor transformation assuming the MSW+LMA solution, as shown in Ref.~\cite{Agostini:2018uly}.

The electron neutrino survival probabilities $\bar{P}_{ee}$ averaged over each energy range of this analysis and calculated with the equations reported in the Appendix, are $\bar{P}_{ee}$($^8$B$_{HER}$, 8.8 MeV) = 0.37 $\pm$ 0.08, $\bar{P}_{ee}$($^8$B$_{HER-I}$, 8.0 MeV) = 0.39 $\pm$ 0.09, and $\bar{P}_{ee}$($^8$B$_{HER-II}$, 9.9 MeV) = 0.35 $\pm$ 0.09\footnote{These average energies for each range have been corrected with respect with those published in Ref.~\cite{Agostini:2018uly} without any appreciable impact on the conclusions reported in that paper.}

\section{Search for solar \MakeLowercase{\textit{hep}} neutrinos}
\label{sec:hep}
We performed a search for \textit{hep} neutrinos looking for their elastic scattering on electrons and their neutral current-mediated inelastic scattering on carbon nuclei, $^{12}$C($\nu,\nu'$)$^{12}$C*, where the excited $^{12}$C nucleus in the final state de-excites by emitting a 15.1~MeV $\gamma$-ray.
The \textit{hep} neutrinos are the only neutrinos produced in the solar \textit{pp}-chain yet to be observed.
They are both the least abundant solar neutrinos but because they are the highest energy ones ($<$18.8 MeV), they are the most sensitive to Mikheyev-Smirnov-Wolfenstein conversion making their study of particular interest for neutrino oscillations at the very long baseline.

In 2006, Super-Kamiokande~\cite{hep.SK.2006} and SNO~\cite{hep.SNO.2006} have each obtained upper limits on the \textit{hep} neutrino flux by looking for their elastic scattering on electrons and their charge current interaction with deuterium.
The Borexino target mass is not optimized for a clear detection of neutrino fluxes at the level of $\sim$10$^{3}$ cm$^{-2}$s$^{-1}$, as predicted for \textit{hep} neutrinos in the SSM~\cite{Vinyoles:2016djt}. However, when complementing the neutrino-electron scattering detection channel with neutrino interactions on carbon, Borexino can set a competitive experimental constraint on their interaction rate.  

For this analysis we used the data acquired by both the primary and the FADC DAQ systems, the latter optimized for the acquisition of high-energy events~\cite{Alimonti:2008gc}, following the approach applied in~\cite{hep.Agostini.2017}. Collected data corresponds to 4.766 live years starting on November 2009 when the FADC system was commissioned, until October 2017. The end date was chosen beyond that used for the $^8$B analysis by $\sim$10~months in order to maximize the statistical power of the dataset. The effective exposure is 745 t$\cdot$y during which the primary and the FADC DAQ systems were concurrently operational. 

\begin{figure}
\includegraphics[width=.9\linewidth]{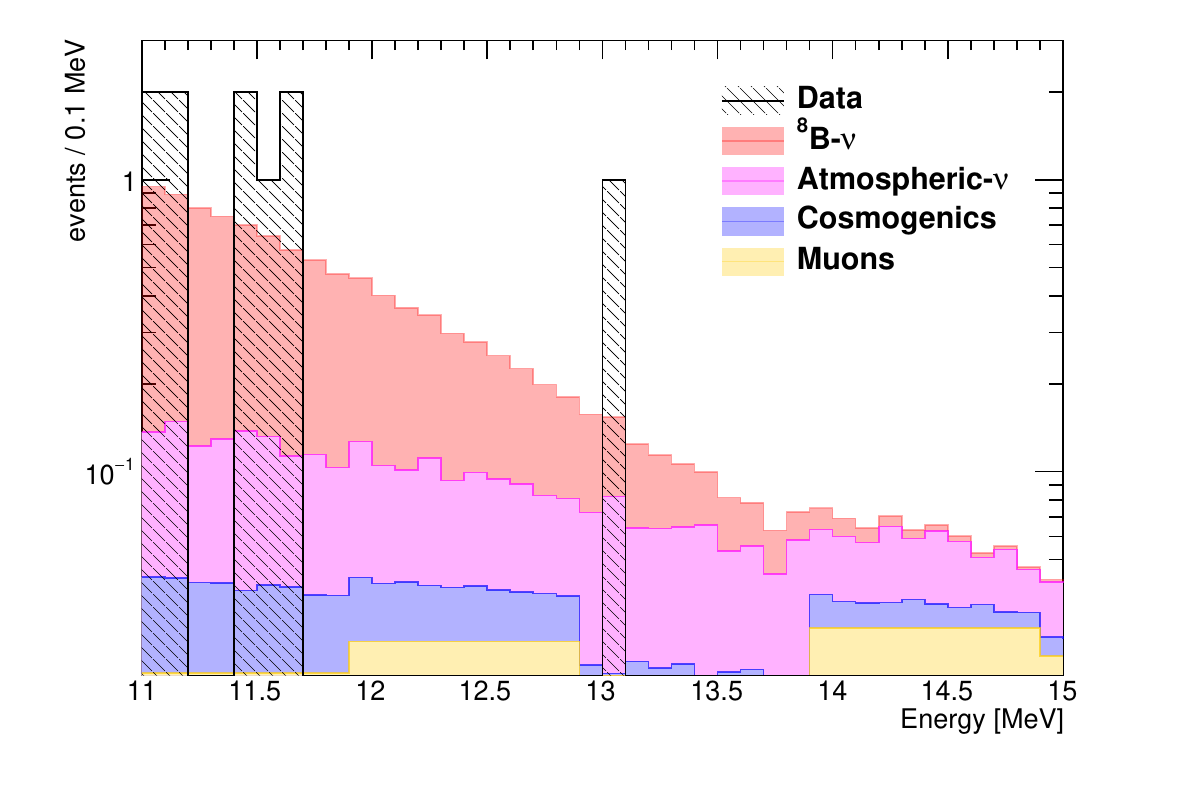}
\caption{\label{fig:hep} FADC energy spectrum of selected events above 11 MeV, compared with the expected background  spectrum.}
\end{figure}

Data are selected by applying the \textit{neutron cut}, the \textit{fast cosmogenics cut}, and the \textit{run start/break cut} described in Section~\ref{sec:data}. We additionally require that the FADC energy of candidate events falls in the [11, 20] MeV energy range and their vertex position is 25~cm away from the nylon vessel, corresponding to a fiducial target mass of $\sim$216 tons. The final sample comprises 10 candidate events surviving all selection cuts, shown in Fig.~\ref{fig:hep}.

The background sources in the \textit{hep} energy range are $^8$B solar neutrinos, atmospheric neutrinos, untagged muons, and long-lived cosmogenic isotopes. The efficiency of the selection cuts for the latter two background components are derived from the analysis described in Section \ref{sec:backgrounds}. 

Untagged buffer muons with low energy deposition in the [11, 20] MeV energy range can mimic the signal of \textit{hep} neutrinos. In contrast to the $^8$B solar neutrino analysis, the number of untagged muons in this energy region of interest is no longer negligible. The FADC DAQ exploits additional algorithms for cosmic muons identification based on a pulse shape analysis. The number of background muon events surviving selection cuts in the $hep$ range is 2.2$\pm$1.5. 

The expected number of events from cosmogenic $^8$B and $^8$Li above 11~MeV is 0.55$\pm$0.20. The contribution from $^{11}$Be in the same range is 0$\substack{+0.0068 \\ -0.0000}$, negligible with respect to the other components.

For the expected number of background $^8$B neutrinos and its uncertainty we use the measurement reported in this paper confined to the \textit{hep} energy range, obtaining to 7.61$\pm$0.54 events. 

The number of background events induced by atmospheric neutrino interactions via charged and neutral currents with protons and carbon nuclei in the scintillator is estimated with the GENIE~\cite{GENIE.2010} Monte Carlo, using the neutrino fluxes from HKKM2014~\cite{PhysRevD.92.023004} above 100 MeV and from FLUKA~\cite{FLUKA.2005} below 100 MeV. The byproducts of $\nu+^{12}$C and $\nu+p$ interactions from GENIE are propagated through the full Borexino Monte Carlo chain. The  number of events induced by atmospheric neutrinos in the \textit{hep} range and exposure after data  selection cuts is 2.4$\pm$1.6. 

In summary, the total number of expected events from backgrounds is 12.8$\pm$2.3, to be compared with the observed 10 events, as shown in figure \ref{fig:hep}.

The analysis reported in~\cite{Agostini:2018uly}, which has the same exposure and data selection methods and uses the Feldman-Cousins approach~\cite{PhysRevD.57.3873}, reports a limit on the number of \emph{hep} neutrino events of 5.56 (90\% C.L.), corresponding to a flux of $<$2.2$\times$10$^{5}$~cm$^{-2}$s$^{-1}$.  In this work we have instead adopted a Profile Likelihood (PL) approach, which accounts for background uncertainties and preserves the consistency with the $^8$B neutrino analysis reported here. 

The PL method is implemented with the HistFactory~\cite{Cranmer:2012sba} package, which also accounts for systematics related to the spectral shape uncertainties.  The new measured limit on the detected number of \emph{hep} neutrino events is 4.37 at 90\%~C.L. The corresponding limit on the \emph{hep} neutrino flux is $<$1.8$\times$10$^{5}$~cm$^{-2}$s$^{-1}$. 

We note that this limit is $\sim$1.2 times stronger than that reported in~\cite{Agostini:2018uly} thanks to a slightly larger contribution of the expected background obtained in this work that originates from the inclusion of the additional background from atmospheric neutrinos. The limit is a factor of $\sim$7 less stringent than reported by SNO ($<2.3\times10^{4}$ cm$^{-2}$s$^{-1}$ (90\% C.L.)~\cite{hep.SNO.2006}) and comparable to that reported by Super-Kamiokande ($<1.5\times10^{5}$ cm$^{-2}$s$^{-1}$ (90\% C.L.)~\cite{hep.SK.2006}).

\section{Conclusions}
\label{sec:conclusions}
In this work, we describe a new analysis of Borexino data that has led to an improved measurement of the $^8$B solar neutrino rate. 
At $\sim$1.5~kt$\cdot$y, this exposure is $\sim$11.5 times larger than that used in a previously-released measurement~\cite{Bellini:2008mr}.  

Key improvements are a lower $^{208}$Tl contamination in the liquid scintillator target achieved after purification, the inclusion of radiogenic neutron captures on detector components to the detector background model, and a tighter constraint applied to the rate of cosmogenic $^{11}$Be. Equally importantly, the analysis rests on a largely improved detector Monte Carlo simulation package, able to model the detector response at a few percent level. These refinements made it possible to extend the analysis to the entire scintillator volume, which reduced the uncertainty on the $^8$B rate from 19\% to 8\%. 

This improved measurement, along with the a tighter constraint on the \textit{hep} neutrino flux, complements the recent work on the simultaneous spectroscopy of $pp$, $^7$Be, and $pep$ solar neutrinos with Borexino Phase-II data, as reported in ref.~\cite{Agostini:2018uly}.

\appendix
\section{$^8B$ rate}
\label{app:Pee}
The expected number of events due to solar neutrino elastic-scattering interactions in Borexino for $^8$B neutrinos can be determined by the following equation:

\begin{widetext}
\begin{eqnarray}
& R_{B8}(T_1,T_2)\;\; &= A\int_0^{E_{max}}dE \,\phi(E) P_{ee}(E) \int_0^{T_{max}(E)} dT\,\frac{d\sigma_e}{dT}(E,T)\,\eta(T;T_1,T_2) + \nonumber \\
&  &+ A\int_0^{E_{max}}dE \,\phi(E) (1-P_{ee}(E)) \int_0^{T_{max}(E)} dT\,\frac{d\sigma_x}{dT}(E,T)\,\eta(T;T_1,T_2) \nonumber \\ 
&  &= A\int_0^{E_{max}}dE \,\phi(E) [P_{ee}(E) \braket{\sigma_e(E)}^{T_2}_{T_1} + (1-P_{ee}(E)) \braket{\sigma_x(E)}^{T_2}_{T_1}]
\label{eq:rate}
\end{eqnarray}
\end{widetext}

\noindent where $E$ is the neutrino energy, $T$ the visible energy of the scattered electron, $\phi$ the neutrino flux at Earth as a function of the energy, $x = {\mu,\tau}$, $P_{ee}$ the electron neutrino survival probability\footnote{The $P_{ee}$ is also function of neutrino oscillation parameters $\Delta m_{12}^2,\theta_{12}, \theta{13}$ in the MSW framework. Yet, here we are interested, as later shown in Eq.~\ref{eq:chi2}, in an effective value determined from experimental data. Therefore, we do not discuss the dependence on these parameters.}, and $A$ = 2.857 a normalization factor with the neutrino flux given in units of $10^9$ cm$^{-2}$ s$^{-1}$, assuming the SSM-HZ model ($\phi= 5.46\times10^6$ cm$^{-2}$ s$^{-1}$), with the cross-section in units of $10^{-45}$ cm$^2$, assuming the electron density equal to $3.307\times10^{23}$ kg$^{-1}$, and with the rate in units of cpd/100 ton. 
In Eq.~\ref{eq:rate} $\eta$ is the detector efficiency function, shown in Fig.~\ref{fig:thresholds}, {\it i.e.} the probability that the scattered electron with visible energy $T$ will be detected in the energy interval of interest $(T_1, T_2)$. 
In Eq.~\ref{eq:rate}, $\braket{\sigma_i(E)}^{T_2}_{T_1}$, with $i={e,x}$ is the electron-neutrino elastic scattering cross section folded on the efficiency function.

\begin{figure}[b]
\includegraphics[width=.9\linewidth]{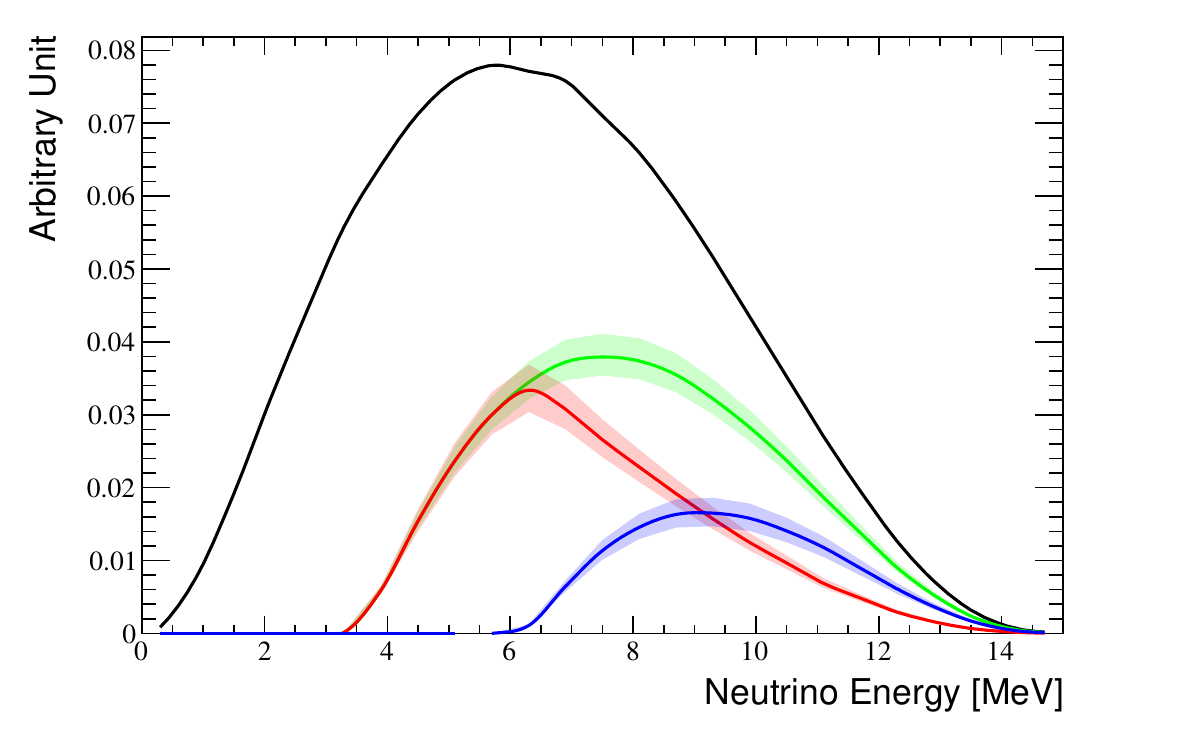}
\vspace{-12pt}\caption{Response functions from Eq.~\ref{eq:rho} for $i=e, x$ corresponding to the HER (green), HER-I (red), and HER-II (blue) energy ranges, respectively. The black line is the $^8$B neutrino spectrum, for reference.}
\label{fig:response}
\end{figure}

%
Another function can be introduced based on Eq.~\ref{eq:rate}, namely:
\begin{equation}
\braket{\rho_i(E)}^{T_2}_{T_1}=\frac{\lambda(E)\,\braket{\sigma_i(E)}^{T_2}_{T_1}}{\int_0^{E_{max}}dE\,\lambda(E)\,\braket{\sigma_i(E)}^{T_2}_{T_1}},
\label{eq:rho}
\end{equation}

\noindent where $\lambda(E)$ is the neutrino energy spectrum normalized to unity and $i ={e,x}$ specifies different neutrino flavors, with $x=\mu,\tau$. Fig.~\ref{fig:response} shows Eq.~\ref{eq:rho}, the fractional neutrino spectrum in the visible energy window of interest. 

It turns out that:

\begin{widetext}
\begin{equation}
R_{B8}(T_1,T_2) = B \int_0^{E_{max}} dE \,\phi(E) [P_{ee}(E) \braket{\rho_e(E)}^{T_2}_{T_1} + (1-P_{ee}(E)) \braket{\rho_x(E)}^{T_2}_{T_1}]
\label{eq:rate2}
\end{equation}
\end{widetext}

\noindent with $B$ = 0.0156 in the case of SSM-HZ. For $\braket{\rho_e(E)}^{16\,MeV}_{2.5\,MeV}$, shown in Fig.~\ref{fig:response}, it turns out that: $\bar{E}_{\nu}=8.1\pm4.7$ MeV.

In order to determine the survival probability from the measurement of the neutrino-electron interaction rate we define:

\begin{widetext}
\begin{equation}
\chi^2(f_{B8},\bar{P}_{ee}) = \left( \frac{R_{B8}(T_1,T_2)-f_{B8}\,B \left[ \bar{P}_{ee} \int_0^{E_{max}} dE\,\braket{\rho_e(E)}^{T_2}_{T_1} + (1-\bar{P}_{ee})\int_0^{E_{max}} dE\,\braket{\rho_e(E)}^{T_2}_{T_1} \right]} {\sigma_{data}} \right)^2 + \left( \frac{1-f_{B8}}{\sigma_{B8}} \right)^2,
\label{eq:chi2}
\end{equation}
\end{widetext}

\noindent where $R_{B8}(T_1,T_2) \pm \sigma_{data}$ is the experimental result in $[T_1, T_2], f_{B8} \pm \sigma_{B8}$ is the corresponding $^8$B solar neutrino flux, normalized to the SSM-HZ. In Eq.~\ref{eq:chi2} $\bar{P}_{ee}$ is an effective electron neutrino survival probability in the energy bin of interest, as shown in Fig.~\ref{fig:response}. 
The second term in the right-hand side of Eq.~\ref{eq:chi2} constrains the neutrino flux to the SSM and removes the degeneracy between the flux and the survival probability. 
By marginalizing Eq.~\ref{eq:chi2} against $\bar{P}_{ee}$ we can determine $\bar{P}_{ee}^{best} \pm \sigma_{ee}$. 
The same argument reported above can be applied to other neutrino sources to determine the $\bar{P}_{ee}$ in different energy regions. 

\begin{acknowledgments}
The Borexino program is made possible by funding from INFN (Italy), NSF (USA), BMBF, DFG, HGF, and MPG (Germany), RFBR (16-29-13014, 17-02-00305, 18-32-20073, 19-02-00097) and RSCF (17-12-01009) (Russia),  NCN (Grant No. UMO 2017/26/M/ST2/00915) (Poland), and   UnivEarthS Labex program of Sorbonne Paris Cit\'e (ANR-10-LABX-0023 and ANR-11-IDEX-0005-02). We acknowledge the generous hospitality and support of the Laboratory Nazionali del Gran Sasso (Italy).
\end{acknowledgments}

\bibliography{biblio.bib}

\end{document}